\documentclass[manuscript]{acmart}
\AtBeginDocument{%
  }

\setcopyright{acmlicensed}
\copyrightyear{2018}
\acmYear{2018}
\acmDOI{XXXXXXX.XXXXXXX}
\acmConference[Conference acronym 'XX]{Make sure to enter the correct
  conference title from your rights confirmation email}{June 03--05,
  2018}{Woodstock, NY}
\acmISBN{978-1-4503-XXXX-X/2018/06}

\usepackage{booktabs}
\usepackage{multirow}
\usepackage{array}

\usepackage{subfig}

\usepackage{enumitem}

\usepackage{fvextra}
\DefineVerbatimEnvironment{PromptBlock}{Verbatim}{
  breaklines=true,
  breakanywhere=true,
  breaksymbolleft={},
  breaksymbolright={},
  breakautoindent=false,
  breakindent=0pt,
  fontsize=\small
}

\usepackage{float}
\usepackage{graphicx}
\usepackage{makecell}
\begin{document}

\title{StoryEcho: A Narrative Mirroring Loop Generative Storytelling System for Picky-Eating Intervention}


\author{Yanuo Zhou}
\authornote{Both authors contributed equally to this research.}
\affiliation{%
  \institution{Department of Computer Science and Technology, Tsinghua University}
  \country{China}
}
\email{zhou-yn24@mails.tsinghua.edu.cn}
\orcid{0009-0009-0652-2196}
\author{Jun Fang}
\authornotemark[1]
\affiliation{%
  \institution{Department of Computer Science and Technology, Tsinghua University}
  \country{China}
}
\email{fangy23@mails.tsinghua.edu.cn}
\orcid{0009-0001-2614-8674}

\author{Yuntao Wang}
\authornote{Corresponding authors.}
\orcid{0000-0002-4249-8893}
\affiliation{%
  \institution{Key Laboratory of Pervasive Computing, Ministry of Education, Department of Computer Science and Technology, Tsinghua University}
  \country{China}
}
\email{yuntaowang@tsinghua.edu.cn}

\author{Yi Wang}
\affiliation{
  \institution{HKUST(GZ)}
  \city{Guangzhou}
  \country{China}
}
\email{ywang183@connect.hkust-gz.edu.cn}
\orcid{0009-0007-0035-8415}

\author{Xiaoyu Xie}
\affiliation{
  \institution{School of Design, Hunan University}
  \city{Changsha}
  \country{China}
}
\email{2864209248@qq.com}
\orcid{0009-0003-5562-8137}

\author{Kaixin Ji}
\orcid{0000-0002-4679-4526}
\affiliation{%
  \institution{Key Laboratory of Pervasive Computing, Ministry of Education, Department of Computer Science and Technology, Tsinghua University}
  \country{China}
}
\email{ji-kx@mail.tsinghua.edu.cn}

\author{Yuanchun Shi}
\orcid{0000-0003-2273-6927}
\affiliation{%
  \institution{Key Laboratory of Pervasive Computing, Ministry of Education, Department of Computer Science and Technology, Tsinghua University}
  \country{China}
}
\affiliation{%
  \institution{Intelligent Computing and Application Laboratory of Qinghai Province, Qinghai University}
  \country{China}
}
\email{shiyc@tsinghua.edu.cn}

\renewcommand{\shortauthors}{Zhou et al.}

\begin{abstract}
Picky eating can limit children's dietary variety and create tension in family feeding routines. Existing food-related technologies often focus on mealtime intervention or standalone educational artifacts, offering limited support for connecting low-pressure narrative engagement with children's real-world food exploration over time. We present StoryEcho, a generative storytelling system centered on a narrative mirroring loop, in which personalized stories model sensory exploration through a persistent counterpart and children's subsequent food encounters are reflected back into narrative feedback and future story development. Informed by a formative study, we designed StoryEcho and evaluated it in a 14-day between-subjects field study with 26 families. Compared with food-personalized generative stories without the narrative mirroring loop, StoryEcho was associated with higher try level, approach, and intake, and lower resistance and caregiver pressure. These findings suggest that narrative mirroring can support children's low-pressure food exploration in family routines, while highlighting design tensions for future generative storytelling interventions.

\end{abstract}




%   \centering
%   \includegraphics[width=\textwidth]{image/teaser.png}
%   \caption{StoryEcho as an interactive generative child-as-actor storytelling loop across family routines. Outside mealtimes, the child creates or revisits a virtual avatar and engages with a personalized story about a target low-preference food, with lightweight sensory and interactive tasks embedded in a low-pressure context. In this story world, the child acts as a virtual protagonist. During everyday eating, the child encounters the target food in real life. After the meal, the parent and child record the child’s trying behavior, which StoryEcho uses to provide positive feedback and generate a behavior-informed story ending. The child can then revisit the updated story, through which real-world trying shapes later narrative development over time.}


\maketitle

\section{Introduction}
% 1页

\begin{figure*}[h]
  \includegraphics[width=\textwidth]{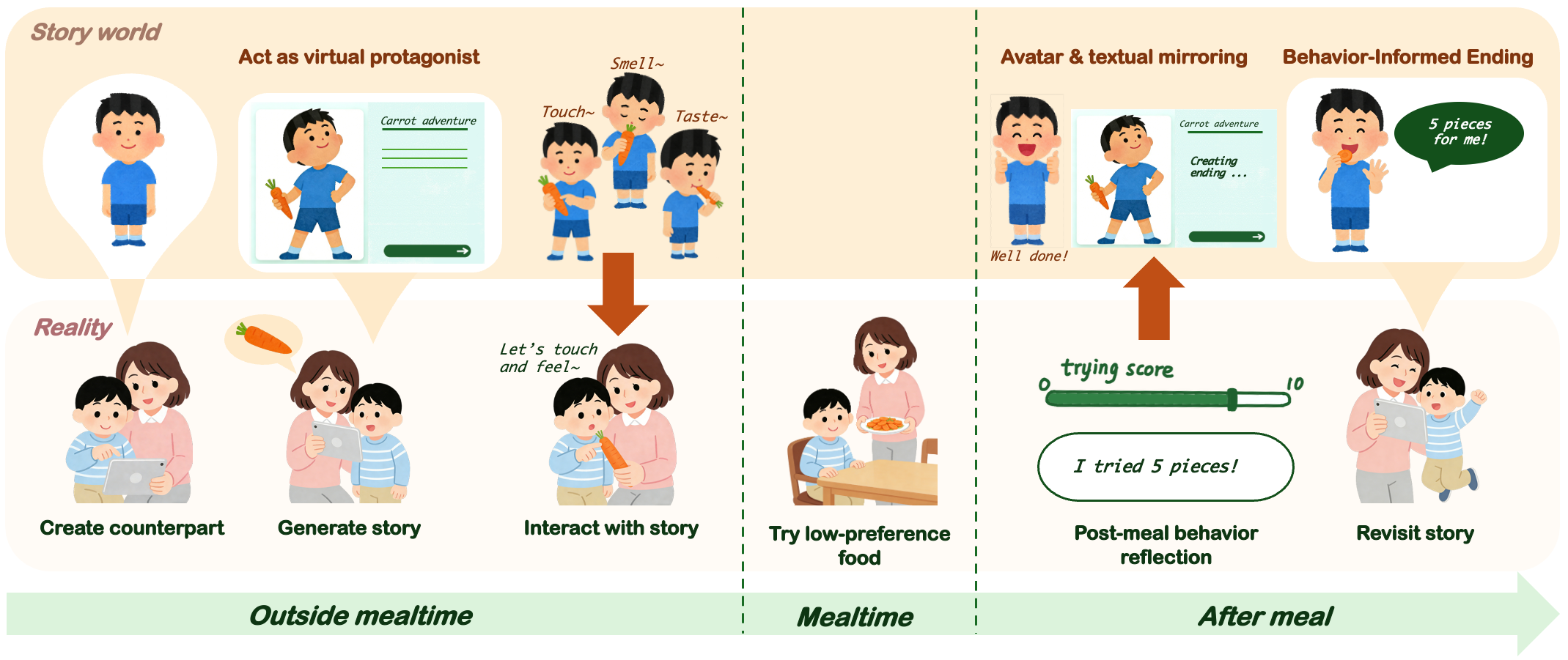}
  \caption{StoryEcho’s narrative mirroring loop connects children’s story-world exploration with their real-world food experiences across everyday family routines. Outside mealtimes, children create a persistent narrative counterpart and engage with personalized stories that model low-pressure sensory exploration of a target low-preference food. During everyday eating, children encounter the food in real life. Afterward, parent--child reflection on the child’s food-exploration behavior is mirrored back into the story through avatar changes, textual responses, and behavior-informed story continuation. In this way, real-world exploration becomes part of subsequent narrative development, sustaining the narrative mirroring loop over repeated use.}
  \label{fig:teaser}
\end{figure*}

% para1: 背景 + 为什么重要
% picky eating 是什么，儿童影响很大。 picky eating intervention for children serves as a behavior change intervention，强调在安全边界xxx下帮助孩童xxxxx

% 挑食的产生原因：和对食物的认知有关。
%Importantly, picky eating and food rejection are shaped not only by sensory preferences or immediate taste experiences, but also by cognitive, social and environmental factors, including how children perceive, mentally represent, categorize, and emotionally respond to foods~\cite{dovey2008food,lafraire2016food}. Thus, addressing picky eating requires gradual changes in how children understand and relate to low-preference foods.

% para2: picky eating intervention 为什么是困难的。现有的一些方式如何开展，不足是什么
% 传统方法：虽多样且具有理论和实证基础，但常以相对独立的策略模块或活动项目存在，缺少将多种策略整合为家庭日常执行流程的结构化工具，可持续性受限。
% HCI研究：大多聚焦餐中行为，缺少认知改善。同时容易打断用餐过程、分散儿童注意力；非餐时干预依赖于特殊设备、材料，对实践有设备要求。

Picky eating is common in early childhood and is typically characterized by reluctance to try unfamiliar foods and rejection of a range of both novel and familiar foods~\cite{dovey2008food,mahmoudizadeh2025interplay,del2024risk}. Such eating patterns can restrict dietary variety, reduce opportunities for repeated food acceptance, and contribute to parental stress and challenging mealtime experiences~\cite{mahmoudizadeh2025interplay,dovey2008food,trofholz2017parents,fries2019parental,del2024risk}. Importantly, food rejection is shaped not only by sensory preferences or immediate taste experiences, but also by how children perceive, categorize, mentally represent, and emotionally respond to foods~\cite{dovey2008food,lafraire2016food}. Thus, supporting children with picky eating involves more than encouraging consumption at a particular meal; it also requires creating opportunities for them to gradually reconsider and approach low-preference foods in more positive ways.

To support such gradual change, prior work on picky-eating intervention has explored repeated exposure, sensory engagement, nutrition education, and parent-mediated support to promote children's food acceptance~\cite{o2012repeated,nekitsing2019taste,weiser2026food,braga2022nutrition,wolstenholme2020childhood,costa2023parental,coe2024exploring,hoppu2015impact,reverdy2008effect}. Technology-mediated approaches have further supported children's eating through sensor-augmented utensils, digital games, and family mealtime technologies~\cite{kadomura2013sensing,jo2020mamas,ganesh2014foodworks,chen2024more}. However, interventions situated directly at mealtimes often center on the immediate act of eating and may also compete for children's attention~\cite{saltzman2019associations,chen2024more}. Approaches beyond meals provide more room for cognitive and perceptual engagement, but often rely on specialized foods, devices, utensils, or activity materials, limiting their integration into everyday family routines~\cite{wang2025play,cai2024see}. Together, these limitations point to a need for lightweight, routine-compatible ways to support children's repeated engagement with low-preference foods beyond mealtimes.

One promising direction is to engage children with low-preference foods through familiar, low-pressure activities outside mealtimes. Storytelling is well suited to this goal because it provides an imaginative space in which children can reinterpret foods without an immediate demand to eat~\cite{garzotto2010interactive,zahedi2025impact,heath2014let}. Yet existing food-related narratives are largely fixed and one-directional: stories may shape how children perceive food, but children's subsequent real-world encounters rarely influence how the narrative develops. LLM-based generative storytelling creates an opportunity beyond one-off personalization, allowing stories to respond to children's evolving experiences over repeated use. This raises a broader design question: how can generative storytelling establish a recurring connection in which narrative engagement supports real-world food exploration and children's subsequent experiences become consequential to the evolving story?

To investigate this question, we conducted a formative study with six parents and six kindergarten educators to understand how narrative reframing might translate into manageable food exploration and how caregivers could support this process without creating pressure. Informed by these findings, we developed the \textit{narrative mirroring loop}: a recurring story--behavior--story connection in which narrative exploration supports subsequent real-world action and children's lived experiences become consequential to later story development. Conceptually, the loop combines persistent narrative identity across story and real-world encounters, bidirectional story--behavior coupling, and longitudinal narrative consequences. Together, these properties extend storytelling beyond one-way behavioral guidance and conventional personalization by linking real-world experiences back to the narrative.

We instantiate the narrative mirroring loop in StoryEcho, an LLM-based generative storytelling system for picky-eating support. StoryEcho operationalizes the loop through a persistent narrative counterpart, a \textit{story-to-practice bridge}, and \textit{real-to-story mirroring}. The counterpart anchors children's participation across repeated use; the \textit{story-to-practice bridge} models low-pressure sensory exploration and invites corresponding real-world actions, while \textit{real-to-story mirroring} reflects children's subsequent food encounters and reflections through immediate avatar and textual responses and later story development. In a 14-day between-subjects field study with 26 families, caregivers in the StoryEcho condition reported higher child \textit{try level}, \textit{approach}, and \textit{intake}, as well as lower \textit{resistance} and \textit{caregiver pressure}, than those in the food-personalized generative-storytelling control condition.

\begin{itemize}

\item We provide empirically grounded design considerations for low-pressure storytelling interventions that support children's participation in picky-eating contexts, derived from a formative study with parents and kindergarten educators.

\item We contribute narrative mirroring as a design mechanism for behavior-responsive generative storytelling, characterized by persistent narrative identity, bidirectional coupling between narrative and real-world action, and longitudinal narrative consequences. StoryEcho instantiates this mechanism through a persistent narrative counterpart, a story-to-practice bridge, and real-to-story mirroring.

\item Through a 14-day between-subjects field study with 26 families, we provide comparative evidence on caregiver-reported behavioral outcomes and characterize the enactment, routine integration, and repeated-use acceptability of the narrative mirroring loop.

\end{itemize}

\section{Related Work}
% 1页
% 每个related works的最后是要收束到 这个方面我们文章做了什么的

\subsection{Picky Eating Intervention}

Prior research on picky-eating intervention has identified several recurring strategies for improving children's food acceptance, including repeated taste exposure, sensory-based food activities, nutrition education, and parent-mediated feeding support~\cite{o2012repeated,nekitsing2019taste,weiser2026food,braga2022nutrition,wolstenholme2020childhood,costa2023parental,coe2024exploring,hoppu2015impact,reverdy2008effect}. Food-related narrative interventions further suggest that stories and picture books can influence children's food choices and willingness to engage with unfamiliar or less-preferred foods, although these interventions have largely relied on pre-authored or fixed story materials~\cite{zahedi2025impact,heath2014let}.

In HCI, related work has explored technologies for children's eating along two main directions. First, many systems intervene during meals to shape children's immediate eating behavior or support parent--child interaction, including sensor-augmented utensils, digital games, and family mealtime technologies~\cite{kadomura2013sensing,jo2020mamas,ganesh2014foodworks,chen2024more}. Although such systems may shape in-the-moment behavior and interaction, mealtime interventions can compete for children's attention and disrupt family eating routines~\cite{saltzman2019associations,chen2024more}. Second, other HCI approaches support children's cognitive or perceptual engagement with food beyond meals through specialized foods, devices, utensils, or activity materials~\cite{wang2025play,cai2024see}. These designs can help children explore and reinterpret foods through sensory or perceptual engagement, yet their reliance on specialized objects or settings may limit flexible use in everyday family life.

Taken together, prior intervention research highlights the value of gradual food exploration, while existing HCI systems offer limited support for lightweight, adaptive engagement across everyday family routines. Our work addresses this gap by connecting low-pressure generative storytelling with real-world food exploration beyond mealtimes.

\subsection{Interactive Storytelling for Children}

HCI research has long explored storytelling as a child-centered interactive medium. Prior work has moved beyond passive story consumption by combining interactivity, multimedia, and embodied engagement to support children's creativity, learning, and active meaning-making~\cite{garzotto2010interactive}. More recent systems support collaborative story creation, visual expression, and child--AI co-creation, positioning children as active participants in narrative development~\cite{zhang2022storydrawer,cai2025child}. LLM- and generative-AI-powered systems further enable personalized reading, responsive interaction, and dynamic content generation based on children's preferences, conversational input, or family context~\cite{chen2025characterizing,xu2025accompany,chen2025scenic}.

Beyond HCI, storytelling has also been examined as a health-education and behavior-change medium for children and adolescents~\cite{richards2026storytelling}. However, HCI research on interactive and generative storytelling for children has paid comparatively less attention to connecting narrative engagement with everyday health-related behavior. Existing child-facing systems primarily allow children's preferences, conversations, or creative inputs to shape content within the storytelling experience. Related HCI work beyond the child-storytelling context has shown that behavioral and affective data can be represented through personalized narratives for reflection~\cite{peng2018trip}, and that tracked physical activity can shape character progress and unlock subsequent story chapters~\cite{murnane2020designing}. These studies demonstrate the potential of behavior-responsive narrative feedback. Yet less is known about how child-facing generative storytelling can support a recurring bidirectional connection in which a story scaffolds manageable real-world action and children's subsequent lived experiences are reflected through feedback and generative continuation of the evolving story. We conceptualize this recurring story--behavior--story connection as narrative mirroring.

Our work brings this behavior-responsive perspective into interactive storytelling for children through a narrative mirroring loop in which story-based modeling supports real-world food exploration and children's subsequent experiences shape narrative development across everyday family routines.

\section{Formative Study}
\label{sec:formative}

Generative storytelling may offer a low-pressure way for children to encounter and reconsider low-preference foods beyond the immediate demands of eating~\cite{zahedi2025impact}. To translate this potential into an everyday intervention, we conducted a formative study with parents and kindergarten educators to understand their observations and interpretations of how children's low-preference food exploration unfolds in everyday routines, how story-based reframing might be connected to real-world practice, and what kinds of caregiver support would be needed to sustain children's engagement. These insights informed design considerations for a generative storytelling system that connects narrative engagement with manageable food exploration in family life.

\subsection{Methods}

\subsubsection{Participants}

We recruited two stakeholder groups with sustained opportunities to observe food rejection among preschool children and to provide food-related support in everyday settings: parents and kindergarten educators. Parents were eligible if they regularly cared for a child aged 3--6 who exhibited picky eating. Educators were eligible if they had practical or professional experience supporting eating practices, nutrition-related activities, or daily caregiving for children in this age range. Participants were recruited through snowball sampling~\cite{goodman1961snowball} and screened for relevant experience. The final sample consisted of parents ($n = 6, M = 36.67, SD = 2.07$) with preschool children ($n = 6, M = 5.33, SD = 0.82$), and kindergarten educators ($n = 6, M = 31.33, SD = 6.56$). Table~\ref{tab:formative_participants} summarizes participant demographics and relevant caregiving or professional experience.

\begin{table}[b]
\centering
\resizebox{0.8\columnwidth}{!}{%
\begin{tabular}{lccccc|lccc}
\toprule
\multicolumn{6}{c|}{Parent participants} & \multicolumn{4}{c}{Educator participants} \\
\textbf{ID} &
\textbf{Gender} &
\shortstack{\textbf{Age}\\[-2pt]\scriptsize (years old)} &
\shortstack{\textbf{Child}\\\textbf{Gender}} &
\shortstack{\textbf{Child Age}\\[-2pt]\scriptsize (years old)} &
\shortstack{\textbf{Picky eating}\\\textbf{years}} &
\textbf{ID} &
\textbf{Gender} &
\shortstack{\textbf{Age}\\[-2pt]\scriptsize (years old)} &
\shortstack{\textbf{Teaching}\\\textbf{years}} \\
\midrule
P1 & Female & 36 & Male & 5 & 2   & T1 & Female & 26 & 3   \\
P2 & Female & 38 & Male & 5 & 3   & T2 & Female & 39 & 3   \\
P3 & Female & 36 & Male & 6 & 4.5 & T3 & Female & 25 & 2.5 \\
P4 & Female & 34 & Male & 6 & 4   & T4 & Female & 30 & 8   \\
P5 & Female & 36 & Male & 6 & 5   & T5 & Female & 28 & 3   \\
P6 & Female & 40 & Male & 4 & 1   & T6 & Female & 40 & 10  \\
\bottomrule
\end{tabular}
}
\caption{Demographic information of participants in the formative study, including parents of children with picky eating (left) and kindergarten educators with relevant caregiving or teaching experience (right).}
\label{tab:formative_participants}
\end{table}

\subsubsection{Procedure and Data Analysis}
% 访谈步骤、编码步骤

The study received approval from our Institutional Review Board (IRB). We developed tailored semi-structured interview protocols for parents and kindergarten educators under a shared set of research goals. Parent interviews focused on preschool children's recurring food refusals, parents' interpretations of these refusals, prior intervention attempts and child responses, and the perceived feasibility of family-based practices. We also asked about storybooks, playful food-related activities, child participation, and opportunities for technology-supported intervention. Educator interviews addressed comparable issues in classroom and kindergarten settings, including food-education activities, child participation, adult mediation, and the feasibility of repeated implementation. All interviews were conducted online in Mandarin Chinese and lasted approximately 30--60 minutes. With participant consent, the interviews were audio-recorded and transcribed verbatim.

The first two authors (A1 and A2) conducted a bottom-up thematic analysis~\cite{braun2006using} in NVivo. After familiarizing themselves with the transcripts, A1 and A2 independently coded an initial subset, refined the coding structure through discussion, and then applied it across the remaining data. Open coding focused on passages concerning food-related expectations, observable child responses, intervention practices, perceived effects, and practical or emotional trade-offs. Codes were then compared within and across parent and educator accounts. We treated participant descriptions of effectiveness as situated perceptions, retaining counterexamples, limitations, and tensions alongside positively described practices. Through iterative theme refinements and research-group discussions~\cite{o2020intercoder}, we organized the codes into three themes: \textit{Cognitive Construction}, capturing how children form food-related expectations in daily life; \textit{Narrative Enactment}, capturing how new interpretations become self-relevant real-world exploration; and \textit{Experiential Continuity}, capturing how caregiver mediation and subsequent responses help children carry the meaning of an encounter into later ones.

\subsection{Findings}
\label{sec:formative_findings}

We derived four design-relevant findings from parents' and educators' accounts of everyday picky-eating support. Together, these findings suggest what a storytelling-based intervention should support: revising negative food perceptions, turning narrative reframing into real-world exploration, sustaining the meaning of exploration over time, and preserving opportunities for child initiative through caregiver mediation.

\subsubsection{F1. Revising negative food perceptions required approachable sensory engagement before tasting.}

Parents and educators described many food refusals as occurring before tasting, which made direct prompts to ``try a bite'' ineffective. Children often approached foods with expectations accumulated from appearance, smell, anticipated texture, resemblance to previously rejected foods, and earlier bodily experiences. For example, P2 recalled that after the child \textit{``once vomited while eating vegetables, he became very resistant, felt that there was something strange in his mouth, and then instinctively rejected anything green.''} Such expectations could also be reinforced through family language and category-based reasoning. Participants observed that repeated family descriptions such as ``he does not eat this food'' could be echoed in children's own statements about what they did or did not eat (P3, P4, T6). Category-based reasoning also shaped refusal: in P6's case, the child's known peanut allergy led him to avoid foods he broadly categorized as nuts, illustrating how safety-related categories could generalize to other foods.

Because refusals often emerged before actual tasting, participants emphasized helping children re-examine foods through sensory encounters rather than immediately requiring intake. T4 personified black fungus and dressed the character in an attractive costume, after which children described it as beautiful rather than dark or unfamiliar. T5 similarly described letting children peel open foods such as an orange and a pomelo to compare \textit{``how they differ and what they are really like,''} which helped them better understand these foods. For design, this suggests that generative storytelling should not stop at narrative reframing, but should connect it with approachable sensory engagement that lets children revisit low-preference foods through concrete actions before being asked to taste.

\subsubsection{F2. Narrative reframing opened possibilities, but real-world exploration benefited from self-relevant, participatory practice.}

Participants described stories, picture books, personified characters, and food-related knowledge as more approachable than direct advice because they allowed children to reconsider foods without an immediate demand to eat. For example, educator T3 described reframing biting broccoli as \textit{``giving the broccoli a haircut''}, which made the food feel playful rather than threatening. However, participants also noted that narrative understanding did not automatically translate into action. As P4 explained, \textit{``when reading the stories, he knows the idea, but he does not put it into practice.''}

What made reframing actionable was concrete involvement that children could recognize as their own. P1 noted that when a story idea was recalled during an actual food encounter, \textit{``the child feels that it really exists and becomes more willing to believe it.''} Parents and educators further emphasized that even small forms of participation could make later eating feel self-relevant: P2 described that when a child helped \textit{``peel it, or wash it, and participated in it himself''}, dinner became more appealing than when \textit{``mom made it alone.''} T5 similarly observed that children often refused chef-prepared vegetables, but were willing to eat vegetables they had helped prepare. Such participation could \textit{``foster pride and connect a later food encounter to their own contribution''} (T2), which participants perceived as strengthening children's involvement and willingness to try. Thus, participants suggested that narrative reframing should be enacted through manageable actions in which children could actively participate, rather than remain an external message.

\subsubsection{F3. Timely feedback affirmed exploration, but simple rewards were not enough to sustain meaning.}

Participants emphasized that timely feedback could support food exploration when it affirmed children's own attempts. By acknowledging small changes, such as accepting an avoided food on the plate, approaching it, or taking a small bite, feedback made children's progress visible and could foster pride. T6 explained that \textit{``when a child takes even one small bite, we may say, `That was great. You made progress compared with before.' It can work quite well.''}
By contrast, external rewards unrelated to the food encounter offered limited sustained support (T3, P3, P2). Television, play, stickers, or similar incentives could motivate children while they remained attractive, but their effects depended on the reward's appeal rather than on children's connection to their own food exploration.

For design, feedback should therefore acknowledge what children noticed or attempted, connect the response to their own exploratory behavior, and carry that meaning forward into subsequent encounters. Sustained support requires feedback that is timely, specific, varied, and integrated into children's ongoing food exploration rather than closed off as a one-time reward.

\subsubsection{F4. Caregiver mediation could support exploration or turn it into compliance.}

Caregivers played a central role in shaping how children interpreted food encounters. Participants noted that preschool children often relied on adults to explain unfamiliar foods, demonstrate possible actions, and help them name sensations they could not yet articulate. For example, educators T1 and T5 observed that children sometimes became more willing to approach a food after watching trusted adults engage with it. In these cases, caregiver mediation appeared to reduce uncertainty and support children's willingness to explore.

However, adult involvement could become counterproductive when it turned exploration into pressure. Parents described reminding, persuading, or directing children before they had a chance to respond independently, such as saying \textit{``You need to eat this''} (P5). In close family settings, repeated expressions of concern or pressure could prompt children to resist before making their own choices.

For design, caregiver mediation should support children's interpretation of food encounters without taking over the experience. Productive mediation involves explaining, modeling, and helping children reflect on what they noticed, felt, or attempted, while leaving room for their own responses and choices. Otherwise, an activity intended to encourage exploration may instead become an exercise in complying with adult expectations.

\begin{table*}[b]

\centering
\small
\renewcommand{\arraystretch}{1.15}
\setlength{\tabcolsep}{5pt}

\begin{tabular}{
    >{\raggedright\arraybackslash}p{3.8cm}
    >{\raggedright\arraybackslash}p{2.0cm}
    >{\raggedright\arraybackslash}p{7.6cm}
}

\toprule

\textbf{Theme(s)} &
\textbf{Supporting Findings} &
\textbf{Design Consideration} \\

\midrule

\textbf{Narrative Enactment} &
\textbf{F2} &
\textbf{DC1.} Support children's active participation and self-relevance throughout the storytelling experience. \\

\midrule

\textbf{Cognitive Construction}\newline
\textbf{Narrative Enactment} &
\textbf{F1, F2} &
\textbf{DC2.} Bridge narrative reinterpretation and real-world food experience through lightweight sensory practice that children can choose how to undertake. \\

\midrule

\textbf{Experiential Continuity} &
\textbf{F3} &
\textbf{DC3.} Carry real-world food-exploration behaviors forward through responsive, varied, and non-judgmental narrative consequences. \\

\midrule

\textbf{Experiential Continuity} &
\textbf{F4} &
\textbf{DC4.} Structure caregiver involvement as supportive co-use that leaves room for children's choices, participation, and self-expression. \\

\bottomrule

\end{tabular}

\caption{Mapping formative themes and findings to the design considerations. Each mapping represents the primary empirical grounding of a design consideration rather than a direct translation into system features.}

\label{tab:formative_mapping}

\end{table*}

\subsection{Design Considerations}
\label{sec:formative_dcs}
Based on these findings, we derived the following four design considerations (DCs) for a parent--child narrative intervention for picky eating, as summarized in Table~\ref{tab:formative_mapping}.

\textbf{DC1: Support children's active participation and self-relevance throughout the storytelling experience.}
The system should support children's active participation and self-relevance across repeated story and food encounters. This may involve an identifiable and continuing narrative counterpart through which children can see their choices and actions reflected in the evolving story.

\textbf{DC2: Bridge narrative reinterpretation and real-world food experience through lightweight sensory practice that children can choose how to undertake.}
The system should translate story-based engagement into optional actions that fit ordinary family routines, such as looking, smelling, touching, comparing, describing, or tasting outside mealtimes, without making swallowing or finishing food the required endpoint.

\textbf{DC3: Carry real-world food-exploration behaviors forward through responsive, varied, and non-judgmental narrative consequences.}
The system should carry children's food-exploration behaviors forward through immediate and longitudinal narrative responses that are specific to what the child noticed or attempted, varied across use, and free from shame, social comparison, punishment, transactional rewards, or pressure tied to the amount consumed.

\textbf{DC4: Structure caregiver involvement as supportive co-use that leaves room for children's choices, participation, and self-expression.}
The system should distribute caregiver roles across the intervention, enabling caregivers to review generated content, scaffold sensory activities, and support reflection. Across these roles, caregiver involvement should remain facilitative and safety-oriented, leaving room for children's choices and initiative rather than becoming persuasion, compliance monitoring, or performance evaluation.

\section{Design of StoryEcho}
%3页

Guided by the design considerations (in Section~\ref{sec:formative_dcs}), we developed the narrative mirroring loop as the design mechanism. We then implemented this mechanism in StoryEcho, a home-based generative storytelling system for supporting children with picky eating.

\subsection{Narrative Mirroring Loop}
% mechanism的三个主要元素：
% 持续的叙事身份 narrative counterpart (DC1)
% 从叙事到现实的实践转化 (DC1, DC2)
% 从现实回到叙事的镜像转化 (DC1, DC3)
% 另外家长将全程参与、辅助以儿童为中心的系统使用

\begin{figure*}[h]
  \includegraphics[width=\textwidth]{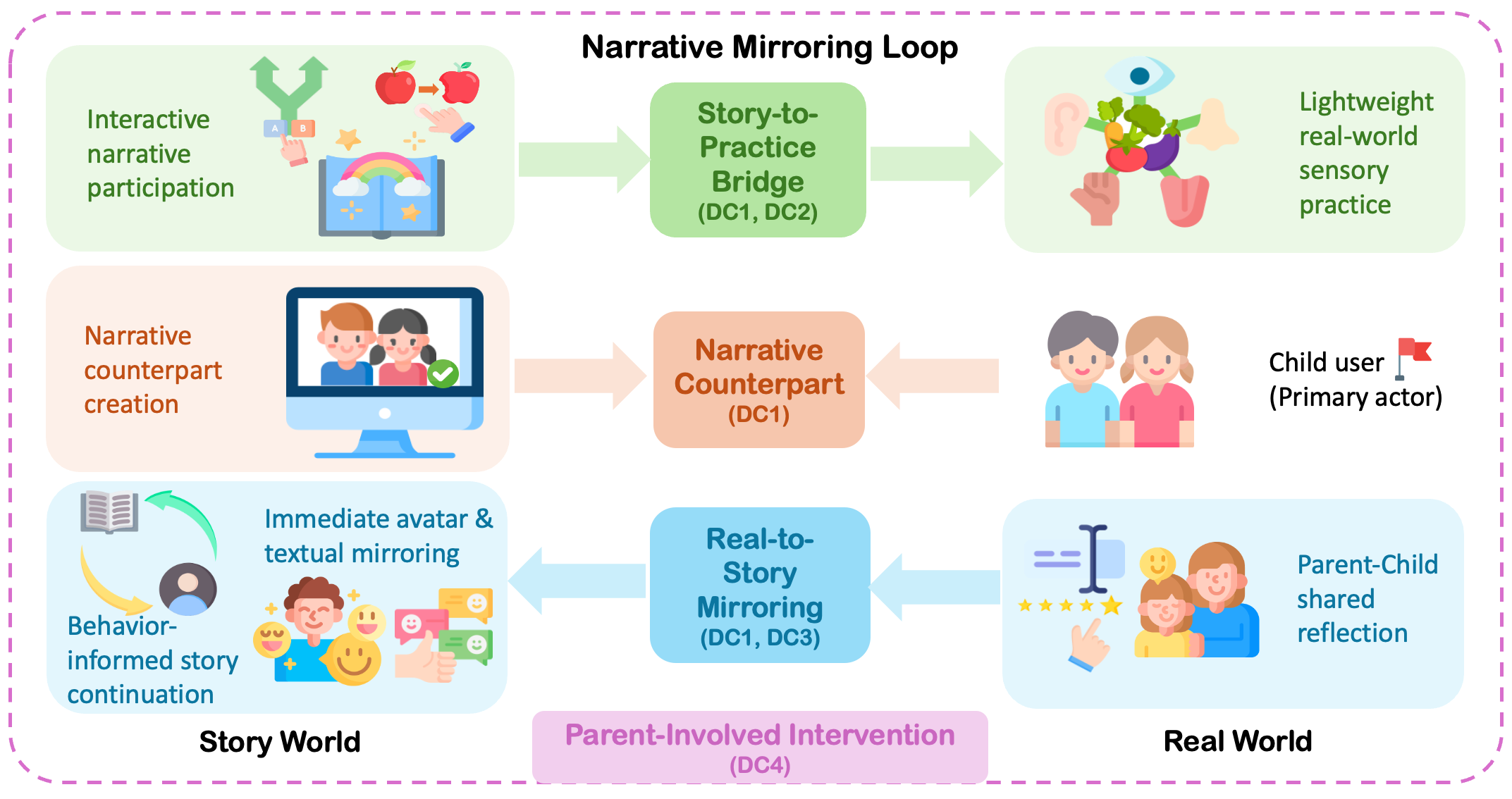}
    \caption{StoryEcho's narrative mirroring loop. The child is positioned as the primary actor, while a persistent narrative counterpart anchors the child's self-representation in the story world (DC1). The upper path shows the story-to-practice bridge, which connects interactive narrative participation to lightweight real-world sensory practice (DC1, DC2). The lower path shows real-to-story mirroring, in which parent--child shared reflection brings the child's real-world exploration back into the story through immediate avatar changes, textual responses, and behavior-informed story continuation (DC1, DC3). The dashed boundary indicates parent-involved intervention across the loop (DC4).}
    \label{fig:narrative_mirroring_loop}
\end{figure*}

The narrative mirroring loop connects children's imagined food encounters in the story world with their real-world exploration of low-preference foods. As a design mechanism, it is characterized by persistent narrative identity, bidirectional story–behavior coupling, and longitudinal narrative consequences. StoryEcho operationalizes these properties through three key elements: a persistent narrative counterpart that sustains children's self-relevance across sessions (DC1), a story-to-practice bridge that guides children from interactive story plots to real-world sensory exploration (DC1, DC2), and real-to-story mirroring that brings children's food-exploration behaviors back into the story through responsive narrative consequences (DC1, DC3). Throughout the loop, caregivers participate as supportive co-users who scaffold children's engagement and sustain participation at home, while children contribute choices, sensory actions, and reflections that shape how the loop unfolds (DC4).
Figure~\ref{fig:narrative_mirroring_loop} illustrates the loop.

\subsubsection{\textbf{Narrative Counterpart}}
% 儿童可以主导创建一个故事中的虚拟形象作为自己在故事中的映射形象。这个形象将作为贯穿整个故事世界的主角，既作为Story-to-Practice Story中的actor，通过“演”出与食物互动的情节，成为孩子的行为榜样，改善孩子对low-preference food的认知，也作为Real-to-Story Mirroring中的actor，表情和反馈话语随孩子的现实尝试产生，并将孩子现实中的尝试行为在故事中“演”出来。

Guided by DC1, StoryEcho uses a persistent narrative counterpart to support children's active participation and self-relevance across sessions. Each child selects and customizes this counterpart as a virtual self-representation in the story world. As a recurring protagonist across sessions, the counterpart provides a stable point of identification and is designed to connect food exploration with children's own participation rather than present it as a generic instructional story.

The narrative counterpart serves as both a role-model and an actor in the narrative mirroring loop. In the story-to-practice bridge, it serves as a role-model by engaging with the target food first, making exploration visible and approachable before the corresponding real-world food encounter. In real-to-story mirroring, it serves as an actor through which children's actual food-exploration behaviors are reflected back into the narrative: its expressions and personalized responses change according to children's attempts, and its later story actions incorporate those attempts. Through this dual role, the counterpart links imagined exploration with real-world practice and provides a consistent point of self-reference across both.

\subsubsection{\textbf{Story-to-Practice Bridge}}
% 包含两个主要部分：
% Interactive Narrative participation 故事中的选择、回应、观察
% Lightweight Sensory practice 现实中的看、闻、摸、比较、描述和尝试
% 故事中的参与：孩子可以通过故事内的互动设计（例如点触互动、情节分支选择活动），主动推动narrative counterpart进行食物感官探索的故事情节。主动控制和参与故事的发展。
% 现实中的实践：故事中设计了一些轻量的感官任务互动（例如感官行为模仿、感官体验描述），让孩子可以在现实中模仿故事中narrative counterpart的食物探索行为。实际参与对食物的感官探索。

Guided by DC1 and DC2, StoryEcho creates a story-to-practice bridge that moves children from narrative reinterpretation to real-world sensory exploration. Rather than presenting the story as passive reading or directly asking children to eat the target food, the bridge first positions children as active participants in the story world and then invites them to carry similar forms of exploration into real life.

\textbf{Interactive narrative participation.} Within the story, children guide their narrative counterpart's food exploration through lightweight interactions such as story-action triggers and plot-branch choices. These interactions allow children to influence how the story unfolds and direct the counterpart to notice, approach, and interact with the target food in a playful context. This frames exploration as something children can help direct through their interactions, rather than a lesson delivered entirely by the system.

\textbf{Lightweight real-world sensory practice.} After story-based exploration, StoryEcho introduces sensory tasks that invite children to explore the actual target food in ways modeled by their counterpart, such as looking, smelling, touching, comparing, describing, or gently tasting. Rather than framing practice as an eating requirement, these tasks offer manageable steps that children can choose whether and how to undertake.

\subsubsection{\textbf{Real-to-Story Mirroring}}
% 包含三点：
% Child-Led Shared Reflection 儿童先表达，家长提供语言和情境支持，共同回顾下反馈给system
% Immediate Avatar and Textual Mirroring 虚拟形象状态和反馈语对具体探索给予即时回应
% Longitudinal Story Continuation 现实的行为改变故事结局和后续叙事连续性
% Child-Led Shared Reflection：孩子表达家长辅助，将现实中的食物尝试行为反馈给系统，增加行为回顾的同时，实现real-to-story的输入
% Immediate Avatar and Textual Mirroring：narrative counterpart的状态根据孩子的现实尝试行为及时发生变化，包含形象表情的变化和与行为直接相关的表扬或鼓励的反馈语。
% Longitudinal Story Continuation：原本的食物故事产生和现实行为直接相关的结尾，孩子的现实行为mirror back为narrative counterpart故事中的行为，同时原始故事和新结尾形成完整故事，成为后续故事的背景，为后续故事构成连续性。

Guided by DC1 and DC3, StoryEcho uses real-to-story mirroring to carry children's food-exploration behaviors back into the story through responsive, varied, and non-judgmental narrative consequences. Rather than treating real-world practice as the endpoint, this mechanism turns children's attempts into material for the ongoing narrative, helping them see their exploration as consequential within the story world while preserving a low-pressure framing.

\textbf{Parent--child shared reflection.} Real-to-story mirroring begins with shared reflection structured around what children did, felt, or noticed during food exploration, while parents provide linguistic and contextual support when needed. This process helps children revisit their attempts without reducing them to success-or-failure judgments, and provides the input through which lived experiences are brought back into the system.

\textbf{Immediate avatar and textual mirroring.} After children's attempts are shared, StoryEcho mirrors them through changes in the narrative counterpart and accompanying textual responses. The counterpart's avatar expressions acknowledge the reported attempts, while textual messages respond to concrete efforts with encouragement. Together, these responses make children's attempts visible in the story world and frame them as meaningful steps rather than offering generic rewards detached from their experience.

\textbf{Behavior-informed story continuation.} Beyond immediate mirroring, children's real-world behaviors also shape story continuation. In tailored endings, their attempts are reflected through the counterpart's actions, connecting the food-related plot with what happened in real life. These endings close the current session while providing narrative context for subsequent stories, carrying children's exploration forward within an evolving story world rather than treating it as a one-time task.

\subsubsection{\textbf{Parent-Involved Intervention}}
% 家长在整个loop中起到支持引导孩子参与、辅助推动loop的作用，保留孩子的使用主体性。由于孩子年龄小，提供内容质量的安全把控和使用上的适当帮助。

Guided by DC4, StoryEcho involves parents as supportive co-users throughout the loop. Because preschool children may need help with reading, operating the system, preparing food, and expressing their experiences, parents scaffold their participation and help sustain the routine at home. This role is intended to support rather than replace children's participation: children's choices, sensory actions, and reported experiences remain the primary inputs shaping how the loop unfolds.

Parents also provide age-appropriate guidance and safety supervision, helping to assess whether generated content and sensory practices are understandable, appropriate, and safe. 
By supporting use without turning exploration into pressure to eat, parent involvement can help sustain the intervention while leaving room for children's choices, participation, and self-expression.

\subsection{System Design and Implementation}

\subsubsection{\textbf{Storytelling Content Generation}}
% 包括四个部分：
% 1、4种story framework设计
% 2、每个episode内容：1. 主要情节：counterpart和食物进行感官探索为主的故事情节，主要包含counterpart的食物感官探索行为、食物感官特征、食物烹饪知识、食物健康知识。 2. 互动：要求故事中必须包含2种故事内互动和2种现实感官任务互动 3. 图片：保持画风一致性，互动会触发图片变化，展现故事中的action发生。
% 3、故事的连续性：1. ending和原故事的情节连续性，将现实行为融入进原情节 2. 通过summary机制将过往食物故事和本次连续，建立食物之间的联系、同时帮助孩子回顾自己的过往尝试行为。
% 4、安全保障：所有文字内容和图片内容设计了严格的安全要求，适合学龄前儿童阅读，要求模型生成过程需按要求审核通过后输出。

\begin{figure*}[h]
    \includegraphics[width=\textwidth]{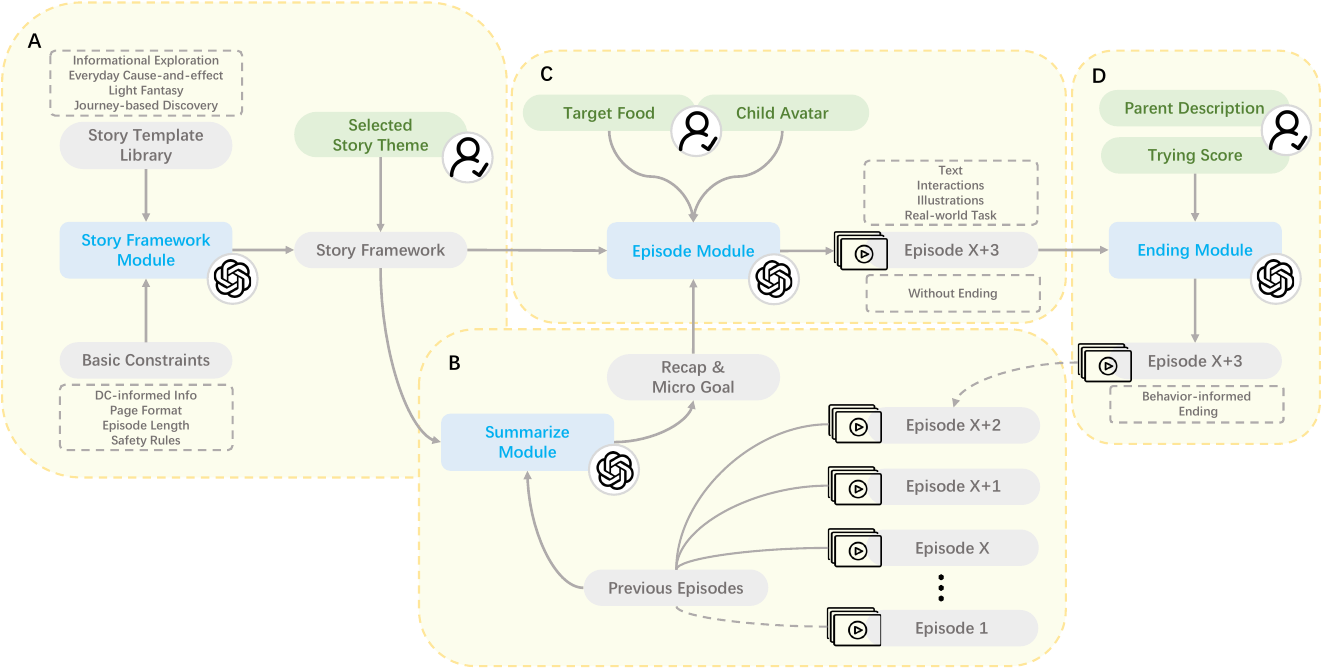}
    \caption{Detailed workflow of StoryEcho's content-generation pipeline. (A) The story framework module combines basic constraints, a story template library, and the selected story theme to produce a stable story framework. (B) Across the continuing picture-book series, the recap module summarizes previous episodes and derives a narrative micro-goal to preserve cross-episode continuity. (C) The episode module combines the story framework, recap and micro-goal, current target food, and child avatar to generate a new episode with text, interactions, and illustrations. (D) After post-meal behavior is recorded, the ending module converts the trying score and behavior description into feedback and a behavior-informed ending that updates the story.}
    \label{fig:generative_storytelling_content_design}
\end{figure*}

StoryEcho generates a continuing series of personalized picture books using story frameworks, episode-level content generation, continuity mechanisms, and content-safety controls. Figure~\ref{fig:generative_storytelling_content_design} illustrates the generation workflow.

\textbf{Story frameworks.} To support repeated yet varied participation, StoryEcho provides four story frameworks from which children select, with caregiver support when needed, based on their interests. Each framework defines the high-level narrative structure, tone, recurring story-world elements, and possible forms of food exploration, while leaving room for episode-level personalization around the child's narrative counterpart and target low-preference food. Rather than generating each story from scratch, these frameworks provide a stable scaffold for narrative coherence and allow the counterpart to remain a recurring actor across episodes.

\textbf{Episode content.} Each generated episode includes story text, interactive elements, and illustrations. Drawing on stakeholder review and prior references, the main episode is structured as a 12-page picture-book story~\cite{shulevitz1997writing}, with 60--80 Chinese characters per page~\cite{montag2015words}. The plot centers on the narrative counterpart's sensory exploration of the target food, incorporating relevant sensory features, simple cooking knowledge, and age-appropriate food-related information. Each episode includes both in-story interactions and real-world sensory practice prompts. Illustrations are generated in a consistent style across pages, and interaction-triggered pages include visual changes that make the counterpart's actions visible and show how children's interactions contribute to the unfolding episode.

\textbf{Story continuity.} StoryEcho maintains continuity at two levels. First, after children report their real-world food-exploration behaviors, a 4-page behavior-informed ending is generated to align with the original plot and complete the current episode. Second, a summary mechanism carries key information from previous episodes into later generation, including prior story events, target foods, and children's exploration attempts. This allows new episodes to revisit past experiences, connect different foods, and build longitudinal progression across the series.

\textbf{Safety control.} Because StoryEcho is designed for preschool children, textual and visual generation follows strict safety requirements. Stories use age-appropriate language, avoid coercive or fear-based persuasion, and frame food exploration in a positive, non-judgmental, and low-pressure manner. Visual content must be child-friendly, stylistically consistent, and free from inappropriate or unsafe elements. Before content is presented to children, StoryEcho applies a prompt-based safety check to review generated story text and visual prompts against these requirements. Parents then review the generated story and can regenerate content they consider unsuitable.

\subsubsection{\textbf{System Usage Pipeline}}

\begin{figure*}[h]
    \includegraphics[width=\textwidth]{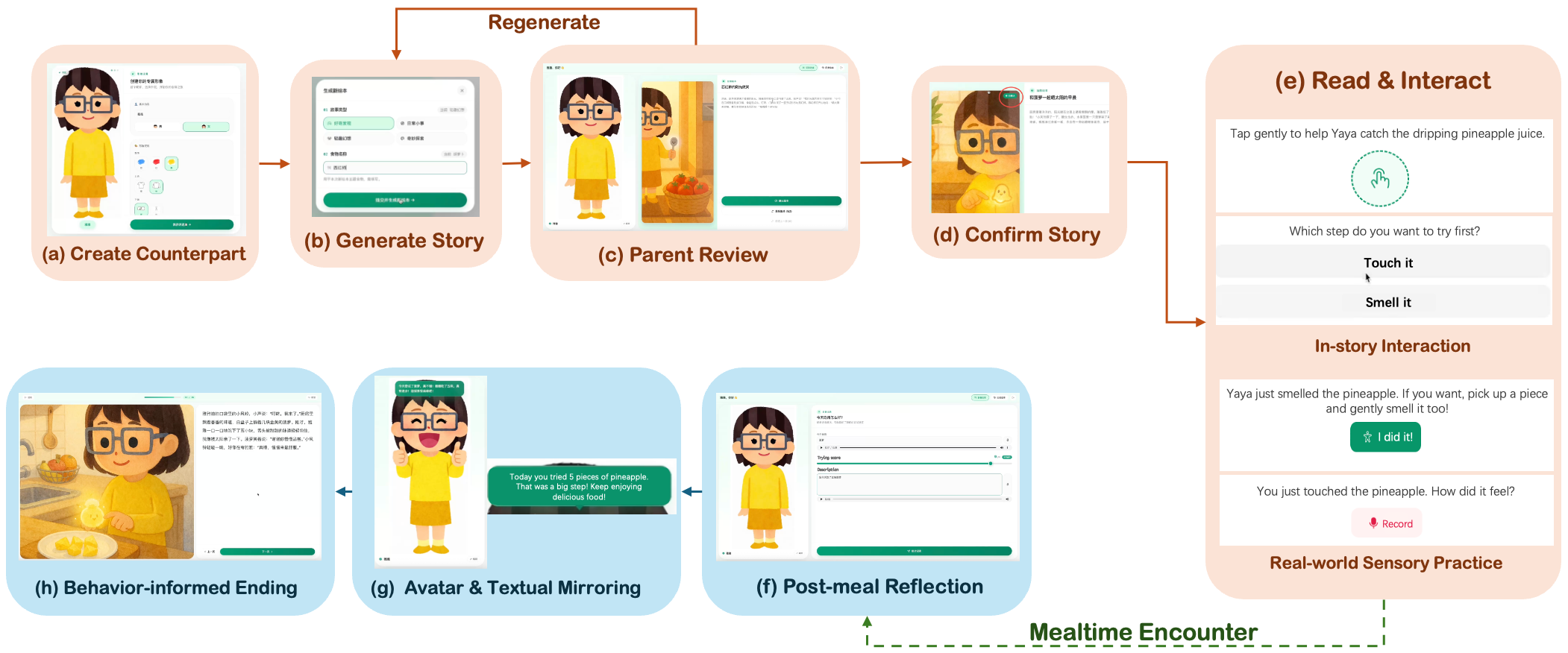}
    \caption{Overview of StoryEcho's usage pipeline. The loop begins with narrative counterpart creation (a) and story generation (b), followed by parental review, optional regeneration, and confirmation (c--d). The child then engages with the confirmed story through in-story interactions and lightweight real-world sensory practices (e). After a mealtime encounter with the target low-preference food, the parent and child complete a post-meal reflection (f), which StoryEcho uses to generate immediate avatar changes, textual responses (g), and a behavior-informed ending (h). The child can then revisit the updated story, completing a narrative mirroring loop that connects outside-mealtime story engagement, real-world food trying, and narrative update.}
  \label{fig:system_loop}
\end{figure*}

StoryEcho's usage pipeline (Figure~\ref{fig:system_loop}) begins with setup: the child selects and customizes a narrative counterpart and chooses a story framework, while the parent specifies a target low-preference food. After StoryEcho generates a personalized picture book, the parent reviews and either confirms or regenerates it. Outside mealtimes, the child reads the story through optional read-aloud support, completes in-story interactions, and responds to lightweight sensory-practice prompts.

Following a mealtime encounter with the target food, the child and parent complete a brief reflection on what the child tried, noticed, or felt. StoryEcho then produces avatar changes, textual responses, and a behavior-informed ending. The child can then revisit the updated story, completing the current narrative mirroring cycle, while a recap of the episode informs subsequent story generation.

\subsubsection{\textbf{Implementation}}

StoryEcho is implemented as a web application with a React 19 and TypeScript frontend and a Python FastAPI backend with SQLite persistence. The backend orchestrates story generation, image creation, prompt-based safety checks, event logging, and feedback APIs. Text-generation tasks, including episode creation, behavior-informed endings, continuity summarization, and textual responses, use \textit{DeepSeek-V4-Pro} Chat Completions. Image generation and editing use \textit{gpt-image-2}. Speech features include text-to-speech via \textit{edge-tts} and audio transcription via \textit{gpt-4o-transcribe-diarize}. A separate Node.js API built with Express and TypeScript handles authentication, user data, and administrative workflows.

\section{User Study}
%2页

We conducted a 14-day between-subjects user study to examine short-term behavioral signals, mechanism enactment, and home-use acceptability of StoryEcho for repeated home use in supporting children with picky eating. The study was guided by one hypothesis and two research questions:

\textbf{H1.} StoryEcho will improve children's willingness to engage with and try low-preference foods relative to food-personalized generative storytelling without the narrative mirroring loop.

\textbf{RQ1.} How were the components of the narrative mirroring loop completed, understood, and perceived to support children's food exploration?

\textbf{RQ2.} How did StoryEcho fit into families' daily routines, and how acceptable was it for repeated and continued home use?

\subsection{Participants}

We recruited 26 families with children aged 3--6 through public social media platforms, including Rednote\footnote{\url{https://www.xiaohongshu.com/}}, with one child participating from each family. Families were eligible if caregivers reported that their child recurrently rejected multiple familiar or unfamiliar foods and was willing to engage with picture book or story-based activities. Because the participating children were preschool-aged, eligibility and food-related information were reported by caregivers. The resulting sample therefore represented children with caregiver-identified picky eating rather than clinically assessed feeding difficulties. Caregivers were also required to nominate 4--5 child-specific low-preference foods that could be safely provided at home during the study. They confirmed that the nominated foods were not associated with any known allergies or medical dietary restrictions for the child. Food-specific eligibility was subsequently determined using the baseline questionnaire, as described below.

The final sample comprised 13 families in the StoryEcho condition ($n = 13$; child age: $M = 4.62$, $SD = 1.19$; caregiver-reported duration of picky eating: $M = 1.62$ years, $SD = 1.02$) and 13 families in the Control condition ($n = 13$; child age: $M = 4.69$, $SD = 1.11$; caregiver-reported duration of picky eating: $M = 2.00$ years, $SD = 1.02$). Table~\ref{tab:user_study_participants} summarizes the child demographics and caregiver-reported duration of picky eating in each condition.

\begin{table}[h]
\centering
\resizebox{\columnwidth}{!}{
\begin{tabular}{cccc|cccc}
\toprule
\multicolumn{4}{c|}{\textbf{StoryEcho Condition}} & \multicolumn{4}{c}{\textbf{Control Condition}} \\
\cmidrule(lr){1-4} \cmidrule(lr){5-8}
\textbf{Participant} & \textbf{Child Gender} & \textbf{Child Age} & \textbf{Picky-eating Years} &
\textbf{Participant} & \textbf{Child Gender} & \textbf{Child Age} & \textbf{Picky-eating Years} \\
\midrule
P1  & Female & 4 & 1 & P1  & Female   & 6 & 2 \\
P2  & Male   & 6 & 4 & P2  & Female   & 6 & 3 \\
P3  & Male   & 6 & 2 & P3  & Male     & 4 & 1 \\
P4  & Male   & 5 & 2 & P4  & Female   & 6 & 2 \\
P5  & Female & 3 & 0.5 & P5  & Female & 3 & 1.5 \\
P6  & Female & 3 & 1 & P6  & Male     & 5 & 4 \\
P7  & Male   & 6 & 1 & P7  & Male     & 4 & 2 \\
P8  & Female & 4 & 2 & P8  & Male     & 5 & 1 \\
P9  & Female & 6 & 1 & P9  & Male     & 4 & 1 \\
P10 & Male   & 4 & 1 & P10 & Female   & 3 & 0.5 \\
P11 & Male   & 3 & 0.5 & P11 & Female & 6 & 3 \\
P12 & Female & 5 & 3 & P12 & Male     & 5 & 3 \\
P13 & Female & 5 & 2 & P13 & Female   & 4 & 2 \\
\bottomrule
\end{tabular}
}
\caption{Demographic information for child participants in the user study, grouped by the StoryEcho and Control conditions.}
\label{tab:user_study_participants}
\end{table}

\subsection{Procedures}
% AB实验为了验证storyEcho narrative mirroring loop机制的作用，因此对照组使用对照系统，去掉storyecho中narrative mirroring loop相关的功能，保留输入食物名称生成绘本的功能，绘本内容生成保持和storyecho的内容一致性：保留4种Story frameworks可选，episode content围绕目标食物的sensory exploration的情节，保留cross-episode story continuity。对照系统UI和实验系统相似，辅助功能如伴读模式均保留。

\subsubsection{Between-subjects study design.}

During the 14-day study, each family was asked to complete seven ``Target Food Opportunities'' (TFOs). We defined a TFO as one study unit centered on a specific low-preference food, involving system use and a subsequent mealtime encounter with that food.

To examine the role of StoryEcho's narrative mirroring loop, we compared the full StoryEcho system with a control version without the loop-specific components. 
Families were randomly assigned at the household level in a 1:1 ratio to the StoryEcho or Control condition. After allocation and before deployment, caregivers completed a food-specific baseline questionnaire for 4--5 nominated foods. Only foods with a caregiver-reported baseline liking score of 4 or below on a 7-point scale (1 = strongly dislikes, 4 = neutral, and 7 = strongly likes) were retained in each child's target-food pool. For each TFO, caregivers randomly selected one target food from the child’s prespecified eligible pool, allowing the same food to be encountered across multiple TFOs.

Families in the StoryEcho condition used the full system with the narrative mirroring loop, including the persistent narrative counterpart, story-to-practice bridge, and real-to-story mirroring. Families in the Control condition used a version without these loop-specific components but with comparable food-personalized picture-book generation. Both systems used the same underlying generative models and comparable prompt templates for picture-book generation. In the control condition, parents could similarly enter a target low-preference food, and children could choose from the same story frameworks as in StoryEcho. The generated episodes depicted sensory encounters with the target food within the story but did not include the interactive or behavior-responsive loop components. The control version also used a similar UI and retained auxiliary features such as the optional read-aloud mode. 
% In both conditions, parents supported children's system use at home.
In both conditions, parents could assist their children in using the system at home. Table~\ref{tab:condition_comparison} summarizes the shared and condition-specific features.

\begin{table}[t]
\centering
\small
\begin{tabular}{lcc}
\toprule
\textbf{Feature} & \textbf{StoryEcho} & \textbf{Control} \\
\midrule
Target-food personalization & \checkmark & \checkmark \\
Story frameworks & \checkmark & \checkmark \\
Comparable picture-book generation & \checkmark & \checkmark \\
Similar interface and read-aloud mode & \checkmark & \checkmark \\
Persistent narrative counterpart & \checkmark & -- \\
In-story interactions & \checkmark & -- \\
Real-world sensory-practice prompts & \checkmark & -- \\
Parent--child post-meal reflection & \checkmark & -- \\
Immediate avatar/textual mirroring & \checkmark & -- \\
Behavior-informed continuation & \checkmark & -- \\
\bottomrule
\end{tabular}
\caption{Features shared across the StoryEcho and Control conditions and loop-specific components included only in StoryEcho.}
\label{tab:condition_comparison}
\end{table}

\subsubsection{Data collection.} 
Because the participating children were aged 3--6 years old, all questionnaire measures were completed by caregivers. Child-related measures were based on caregivers' observations during each TFO, while caregiver-related measures captured caregivers' own experiences and perceptions.

Before deployment, parents in both conditions completed a food-specific baseline questionnaire for each low-preference food. The questionnaire assessed \textit{liking}, \textit{perceived difficulty}, prior \textit{try level}, \textit{resistance}, and the amount of \textit{caregiver prompting} typically required. 
After each TFO, they completed a session-specific questionnaire. The questionnaire recorded the highest \textit{try level} reached by the child, the \textit{independence} of the child in approaching the target food, \textit{willingness}, \textit{intake}, \textit{resistance}, \textit{caregiver prompting}, \textit{caregiver stress}, \textit{system burden}, \textit{child enjoyment}, \textit{joint caregiver--child enjoyment}, and \textit{continued-use intention}. We also recorded the interval between story use and the target-food encounter and the meal context of each TFO.
For StoryEcho TFOs, additional items assessed completion and perception of the narrative mirroring loop, attention and responses to narrative changes, recognition of the relationship between real-world behavior and story development, expectations for subsequent story changes, and child-initiated engagement. StoryEcho-specific items also assessed caregiver review, inappropriate generated content, child reactions to feedback, and caregiver-perceived feedback pressure. All questionnaires were administered in Mandarin Chinese. 
The exact item wording, response options, and scoring are provided in Appendix~\ref{app:questionnaire_measures}.

After the deployment, parents in the StoryEcho condition completed the System Usability Scale (SUS)~\cite{brooke1996sus} and participated in semi-structured interviews. The interviews examined the perceived role of the narrative mirroring loop, the integration of StoryEcho into family routines, repeated and continued use, practical burden, and safety-related experiences.

\subsubsection{Ethics and consent.} The study protocol was reviewed and approved by the institutional review board (IRB). Before participation, parents received detailed information about the study purpose, procedure, data collection, potential risks, and privacy protection. Written informed consent was obtained from all participating parents, and children participated with parental permission. Families were informed that participation was voluntary and that they could withdraw from the study at any time.

\subsection{Analysis Approach}

All questionnaire-based outcomes were caregiver-reported. The same caregiver-reporting procedure and parallel outcome measures were used in both conditions. Accordingly, the condition analyses estimate differences in caregiver-reported outcomes rather than independently observed changes in children's eating behavior.

All 26 families completed the seven protocol TFOs, yielding 182 TFOs with 91 TFOs in each condition. Because the outcomes were ordinal and repeated observations were nested within families, we fitted cumulative link mixed models (CLMMs) with a logit link, flexible thresholds, and a participant-level random intercept. The Control condition was the reference level, and continuous covariates were mean-centered. 
% Baseline comparability was assessed descriptively using participant-equal standardized mean differences (SMD).
We then examined whether the two conditions were comparable before the intervention by comparing participant-equal standardized mean differences (SMD).
Despite random assignment, the small sample resulted in chance differences in the target foods assigned to each condition.  At baseline, children in the StoryEcho condition had higher liking scores ($SMD = 0.85$), lower perceived difficulty ($SMD = -0.60$), and higher try levels ($SMD = 0.45$) for their assigned target foods than those in the Control condition. Baseline resistance and caregiver prompting were more similar between conditions ($|SMD| = 0.20$). To account for these imbalances, we matched each TFO to the baseline assessment of the same child--food pair and included the corresponding baseline measure as an outcome-specific covariate in each inferential model.

\subsubsection{H1}
% H1 is primarily evaluated by \textit{try level}, that
The primary outcome for H1 was \textit{try level}, measured on a 10-point ordered scale ranging from complete refusal to voluntary, near-normal consumption of the target food. We examined four secondary outcomes, including \textit{approach}, \textit{intake}, \textit{resistance}, and \textit{caregiver pressure}. \textit{Approach} combined child \textit{independence} and \textit{willingness}, whereas \textit{caregiver pressure} combined \textit{caregiver prompting} and \textit{caregiver stress}. Both pairs showed strong session-level consistency (\textit{approach}: $\alpha = .881$, Spearman's $\rho = .796$, \textit{caregiver pressure}: $\alpha = .913$, $\rho = .839$), which remained acceptable after within-participant centering ($\alpha = .780$ and $.840$). We therefore used the item means as prespecified composite outcomes and retained the component items for auditability.

The primary model included \textit{condition}, the food-specific baseline \textit{try level}, and \textit{TFO position} within the opportunity protocol. Each secondary model adjusted for the most closely corresponding food-specific baseline measure. We applied Holm correction across the four secondary-outcome condition tests. Exploratory repeated-exposure analyses examined both overall TFO order and repeated encounters with the same target food.

\subsubsection{RQ1}
\label{chap:evaluation_RQ1}
We organized the questionnaire evidence into three analytical dimensions: \textit{loop enactment}, indicating whether families completed the intended workflow, \textit{loop salience, response, and recognition}, indicating whether children noticed, responded to, or recognized the relationship between real-world activity and narrative change, and \textit{child-initiated uptake}, indicating relevant actions or expressions that occurred beyond the required workflow.
For post-deployment interviews, we analyzed \textit{component-level perceived helpfulness}, focusing on how caregivers understood each loop component's role in supporting children's food exploration.

\subsubsection{RQ2}
To characterize routine integration, we first summarized session-level usage timing, including the interval between StoryEcho use and target-food encounters (TFOs) and the meal context of TFOs. We then analyzed four session-level outcomes: \textit{system burden}, \textit{child enjoyment}, \textit{joint caregiver--child enjoyment}, and \textit{continued-use intention}. Each outcome was modeled as a function of \textit{condition} and \textit{TFO position}, with a participant-level random intercept. No baseline covariates were included. We applied Holm correction across these four condition tests. SUS scores were summarized descriptively for the StoryEcho condition. For post-deployment interviews, we analyzed routine integration, repeated-use acceptability and continued-use interest, use burden, and caregiver perceptions of safety and emotional pressure.

\section{Results}
% 把每个段落的论点（结论）提前到第一句！加粗

% \subsection{H1: StoryEcho Improved Children's Willingness to Engage with and Try Low-Preference Foods}
\subsection{H1: StoryEcho Was Associated with Greater Willingness to Engage with and Try Low-Preference Foods}

% StoryEcho will improve children's willingness to engage with and try low-preference foods relative to food-personalized generative storytelling without the narrative mirroring loop.
% 主要用定量分析表现，可以辅以一些定性分析：泛化到其他食物上的效果

During the TFO protocol, children in the StoryEcho condition showed higher willingness and independence in approaching and trying low-preference foods, lower resistance, lower caregiver pressure, and higher intake than children receiving food-personalized generative stories without the narrative mirroring loop.

Children in the StoryEcho condition ($M = 7.46$, $SD = 2.03$) reached higher caregiver-reported \textit{try levels} than those in the Control condition ($M = 4.76$, $SD = 2.35$). In the adjusted CLMM, StoryEcho was associated with substantially greater odds of reaching a higher \textit{try-level} category ($OR = 18.28$, $95\%~CI = [5.15, 64.93]$, $p < .001$). 

The four secondary outcomes showed a consistent pattern favoring StoryEcho and the condition effects remained statistically supported after Holm correction. Children in the StoryEcho condition ($M = 4.68$, $SD = 1.38$) showed greater \textit{approach} than those in the Control condition ($M = 3.22$, $SD = 1.38$; $OR = 13.68$; $95\%~CI = [3.71, 50.44]$, $p_{\mathrm{Holm}} < .001$), reflecting higher independence and willingness when engaging with the target food. Children in the StoryEcho condition also reached higher \textit{intake} levels (StoryEcho: $M = 2.65$, $SD = 1.74$; Control: $M = 1.43$, $SD = 1.52$; $OR = 6.78$, $95\%~CI = [1.43, 32.22]$, $p_{\mathrm{Holm}} = .016$). They also showed lower \textit{resistance} (StoryEcho: $M = 2.59$, $SD = 1.32$; Control: $M = 3.98$, $SD = 1.37$; $OR = 0.068$, $95\%~CI = [0.018, 0.252]$, $p_{\mathrm{Holm}} < .001$). At the caregiver level, StoryEcho was associated with lower \textit{caregiver pressure} (StoryEcho: $M = 3.09$, $SD = 1.38$; Control: $M = 4.69$, $SD = 1.30$; $OR = 0.042$, $95\%~CI = [0.010, 0.179]$, $p_{\mathrm{Holm}} < .001$), indicating less prompting and lower experienced stress around the target-food encounter. 

\begin{figure*}[h]
    \centering
    \includegraphics[width=\linewidth]{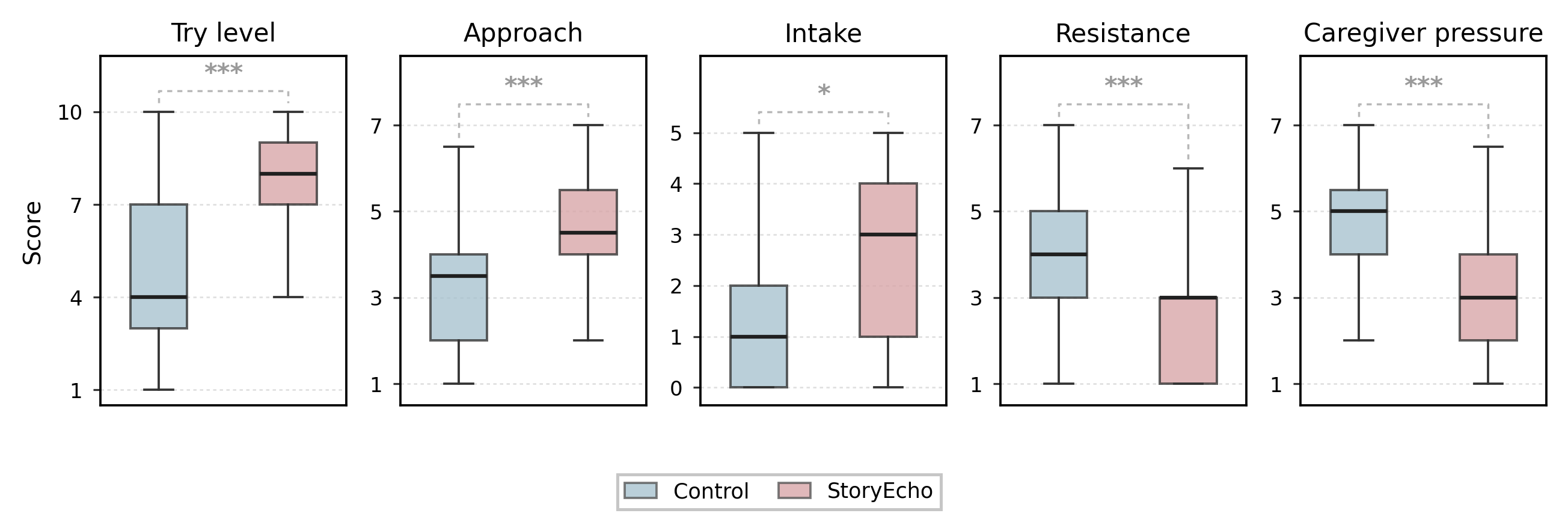}
    \caption{Box plots of caregiver-reported TFO-level outcomes for \textit{try level}, \textit{approach}, \textit{intake}, \textit{resistance}, and \textit{caregiver pressure} across the StoryEcho and Control conditions. Asterisks denote the adjusted condition effects estimated using CLMM. The marker for \textit{try level} reflects the prespecified primary test, whereas the markers for the four secondary outcomes reflect Holm-adjusted $p$ values ($*p < .05$, $**p < .01$, and $***p < .001$).}
    \label{fig:H1_quantitative}
\end{figure*}

Exploratory analyses further suggested that \textit{try level} increased over the seven TFOs in both conditions, with a steeper estimated increase in StoryEcho. For each additional TFO, the odds of reaching a higher \textit{try-level} category increased by a factor of 1.23 in Control ($95\%~CI = [1.02, 1.48]$, $p = .030$) and 1.66 in StoryEcho ($95\%~CI = [1.34, 2.05]$, $p < .001$). The condition-by-TFO-position interaction was positive ($OR = 1.35$, $95\%~CI = [1.03, 1.78]$, $p_{\mathrm{Holm}} = .032$). A separate analysis restricted to repeated encounters with the same food ($N = 24$ families, $n = 116$ TFOs) showed a similar, but less precisely estimated, interaction ($OR = 6.76$, $95\%~CI = [1.53, 29.91]$, $p_{\mathrm{Holm}} = .023$). These exploratory results indicate comparatively faster improvement during the observed TFO protocol.

\subsection{RQ1: Enactment and Perceived Role of the Narrative Mirroring Loop}
% How were the components of the narrative mirroring loop completed, understood, and perceived to support children's food exploration?
% 1. 定量分析从参与度和反馈程度出发分析，分三个层级：参与了环节、注意到且对环节有一定的回应、主动发起环节或者更强烈的反应(比如主动要求下一次或者明确表示行为故事关联)。得到的定量结论更偏向于看narrative mirroring loop整体上的效果，大部分session中都是被回应的，且有一定比例有更主动更强烈的反馈。 
% 2. 定性分析从各个环节出发分析，分析家长对机制中环节作用的理解，放上那个helpful和top3的表格和每个环节的quote，做一个环节之间作用的横向分析。

% 这里放定量分析 enactment 参与度，三个层次

% To further understand how families perceived StoryEcho's narrative mirroring loop, we analyzed parent interviews around five loop components: the narrative counterpart, interactive story participation and sensory practice, parent-child shared reflection, immediate avatar and textual mirroring, and behavior-informed story continuation. 
% All five components were mentioned as helpful by at least some families. Below, we describe how each component was perceived to support children's food exploration, then compare their relative salience across families.

To examine how the narrative mirroring loop was completed, understood, and perceived to support food exploration, we integrated parent-reported indicators from the 91 protocol-matched StoryEcho TFOs with the 13 post-study parent interviews. We focus on five loop components: the narrative counterpart, interactive story participation and sensory practice, parent-child shared reflection, immediate avatar and textual mirroring, and behavior-informed story continuation. We first characterize how consistently and how deeply the loop was enacted across repeated TFOs. We then analyze the interview responses to explain how families interpreted the role of each component and the conditions under which the intended connections became more or less apparent.

\textbf{The loop was consistently enacted and generally perceptible, while child-initiated uptake emerged gradually.}
Table~\ref{tab:rq1_loop_evidence} summarizes caregiver-reported indicators across three analytical dimensions defined in Section~\ref{chap:evaluation_RQ1}.

The loop was consistently enacted, and children generally recognized and responded to its key changes. Completion of the five main loop stages ranged from 82.4\% to 98.9\%, and each stage occurred at least once in all 13 families. Children also commonly noticed and responded to the loop. Loop recognition and response indicators were reported in 83.5\%--94.5\% of TFOs and in every family, including attention to food, recognition of avatar/story changes and behavior--story links, positive responses to feedback and expectation of story continuation.

The loop moved further toward child-initiated uptake across repeated TFOs. Stronger indicators of child initiative occurred less often within individual TFOs but were broadly distributed across families. In 11 of 13 StoryEcho families, children expressed clear anticipation or spontaneously mentioned a subsequent story change, suggesting more active engagement with the whole loop.
To examine how concrete child-initiated behaviors unfolded across repeated use, we calculated a descriptive count of four child-initiated behaviors in each TFO: \textit{spontaneously mentioning food or story features, initiating a sensory action, actively describing the encounter during shared reflection,} and \textit{spontaneously commenting on a story or avatar change}. 
The mean count showed a descriptive increase across the seven TFO positions ($M = 0.54$, $0.61$, $1.15$, $1.31$, $1.31$, $1.31$, and $1.39$, respectively). This pattern suggests that child-initiated uptake became more common after repeated sessions and remained at a higher level thereafter.

\begin{table*}[b]
\centering
\footnotesize
\renewcommand{\arraystretch}{1.13}
\setlength{\tabcolsep}{4pt}
\begin{tabular}{
    >{\raggedright\arraybackslash}p{3.2cm}
    >{\raggedright\arraybackslash}p{6cm}
    >{\centering\arraybackslash}p{2.1cm}
    >{\centering\arraybackslash}p{2.7cm}
}
\toprule
\textbf{Analytical dimension} &
\textbf{Caregiver-reported indicator} &
\shortstack{\textbf{TFOs}\\\textbf{$n/91$ (\%)}} &
\shortstack{\textbf{Families with $\geq 1$}\\\textbf{occurrence, $n/13$ (\%)}} \\
\midrule

\multirow[t]{5}{3.2cm}{\textit{Loop enactment}}
& Story interaction basically completed
& 75 (82.4\%)
& 13 (100\%) \\

& At least part of the sensory task completed
& 88 (96.7\%)
& 13 (100\%) \\

& Child substantively participated in \textit{parent--child shared reflection}
& 85 (93.4\%)
& 13 (100\%) \\

& \textit{Avatar and textual feedback} viewed or heard
& 90 (98.9\%)
& 13 (100\%) \\

& \textit{Behavior-informed ending} read
& 90 (98.9\%)
& 13 (100\%) \\

\midrule

\multirow[t]{5}{3.2cm}{\textit{Loop recognition and response}}
& Story elicited food-related attention
& 86 (94.5\%)
& 13 (100\%) \\

& Child noticed an avatar or story change
& 83 (91.2\%)
& 13 (100\%) \\

& At least possible recognition of the behavior--story relationship
& 82 (90.1\%)
& 13 (100\%) \\

& Child responded positively to feedback
& 76 (83.5\%)
& 13 (100\%) \\

& At least some expectation of a subsequent story change
& 86 (94.5\%)
& 13 (100\%) \\

\midrule

\multirow[t]{6}{3.2cm}{\textit{Child-initiated uptake}}
& Child expressed clear anticipation or spontaneously mentioned a subsequent story change
& 48 (52.7\%)
& 11 (84.6\%) \\

& Child spontaneously mentioned food or story features
& 35 (38.5\%)
& 10 (76.9\%) \\

& Child initiated a sensory action
& 25 (27.5\%)
& 9 (69.2\%) \\

& Child actively described the encounter while the caregiver recorded
& 24 (26.4\%)
& 6 (46.2\%) \\

& Child spontaneously commented on a story or avatar change
& 15 (16.5\%)
& 6 (46.2\%) \\

% & Child explicitly articulated or demonstrated the behavior--story relationship
% & 13 (14.3\%)
% & 5 (38.5\%) \\

\bottomrule
\end{tabular}
\caption{Caregiver-reported evidence of narrative mirroring loop enactment, salience, response, recognition, and child-initiated uptake across the 91 StoryEcho TFOs. These descriptive frequencies characterize the depth and distribution of loop enactment rather than the independent effectiveness of individual components.}
\label{tab:rq1_loop_evidence}
\end{table*}

These TFO-level indicators characterize the breadth and depth of mechanism enactment. We next use the interviews to explain how caregivers interpreted the role of each component, why some components were more readily perceived than others, and where the intended narrative connection remained weak or required caregiver support.

\textbf{Parent interviews suggested that the loop components supported children's food exploration through complementary forms of identification, interaction, reflection, feedback, and continuation.}
The narrative counterpart helped children identify with the story, first through self-selection and then through a sense that the avatar mirrored their own actions (P6, P8, P13); as P13 explained, \textit{``he felt he could control the little character: whatever he did, the character would do as well.''} 

The story-to-practice bridge, including interactive narrative participation and real-world sensory practice, reduced negative feelings and increased curiosity by turning food exploration into gradual, hands-on steps rather than direct eating (P2, P5, P13); P2 noted that her child, who had long been afraid of eggplant, touched it after the story prompted a sensory task: \textit{``I asked him to gently touch it, and he did. That was a big step.''} 

Within real-to-story mirroring, the three components played distinct roles. Parent-child shared reflection made progress recordable and was often caregiver-scaffolded, as preschool children typically verbalized their feelings while parents helped enter them (P1, P3), though some children became more active over time; P10 described that by later sessions, \textit{``he basically did the recording himself.''} Immediate avatar and textual mirroring made children's attempts visible while keeping feedback encouraging rather than judgmental (P3, P6, P8, P12), with P3's child saying, \textit{``If I try a few more bites tomorrow, I want to see whether it will show a different expression.''} Finally, behavior-informed story continuation brought children's real-world attempts back into the narrative, making the real-to-story link vivid and personally meaningful (P2, P6, P7); P6 explained that it made them \textit{``feel that the child's changes were recorded in the picture book, and that the story was rewritten according to the child's eating habits.''}

\textbf{Relative salience of loop components.}
We descriptively summarized family-level mentions and top-component rankings for each loop component, counting repeated mentions by the same family once. As shown in Table~\ref{tab:rq1_component_mentions}, immediate avatar and textual mirroring was the most consistently recognized component: all 13 families described it as helpful, and 8 ranked it among the most helpful components. The story-to-practice bridge and narrative counterpart were also recognized by most families, indicating the salience of both personalized story identification and low-pressure sensory exploration. Parent-child shared reflection was less often singled out, partly because it was frequently parent-mediated rather than child-led. Behavior-informed story continuation was helpful for most families, but depended on whether children recognized the new ending as reflecting their own recent experience; when this connection was unclear, parents found it less impactful (P1, P8).

\begin{table}[h]
\centering
\begin{tabular}{p{0.23\linewidth}p{0.36\linewidth}cc}
\toprule
\textbf{Design mechanism} & \textbf{Loop component} & \textbf{Helpful} & \textbf{Top 3 Helpful} \\
\midrule
Narrative Counterpart & Narrative Counterpart & 10/13 (76.9\%) & 6/13 (46.2\%) \\

Story-to-Practice Bridge & Interactive Participation + Sensory Practice  & 11/13 (84.6\%) & 5/13 (38.5\%) \\

Real-to-Story Mirroring & Parent-Child Shared Reflection & 4/13 (30.8\%) & 2/13 (15.4\%) \\

Real-to-Story Mirroring & Immediate Avatar and Textual Mirroring & 13/13 (100\%) & 8/13 (61.5\%) \\

Real-to-Story Mirroring & Behavior-Informed Story Continuation & 8/13 (61.5\%) & 2/13 (15.4\%) \\
\bottomrule
\end{tabular}
\caption{Family-level mentions of perceived helpfulness across StoryEcho's narrative mirroring components. Counts are descriptive; repeated mentions by the same family were counted once.}

\label{tab:rq1_component_mentions}
\end{table}

\subsection{RQ2: StoryEcho Fit Into Home Routines and Showed Potential for Repeated and Continued Use}
% How did StoryEcho fit into families' daily routines, and how acceptable was it for repeated and continued home use?
% 1. 日常使用时机和融入rountine：定量放一个统计TFO使用时间的图，然后写定性quote里几种常见的融入日常时间的方式。
% 2. 重复使用接纳度&持续使用意愿：先放定量分析里的child enjoyment、joint caregiver-child enjoyment 和下次继续使用意愿 这3条数据的数据分析，定量说明整体使用愉悦程度和意愿相比对照组都是高的。然后定性分析按照实验内 重复使用接纳度，持续使用意愿 这两个分段说明。
% 3. 使用负担：先放定量分析里的caregiver burden和SUS分析，定量说明整体burben是少的。然后定性说明主要负担来源于原型实现的细节，而非 StoryEcho 的概念流程。
% 4. 安全和压力相关：先放定量安全性里的统计数据，表示没有什么不安全内容 + 儿童压力很少。然后放定性分析说caregiver安全性审核的作用，优点以及一些不足

To understand how StoryEcho fit into families' daily routines and whether it was acceptable for repeated and continued home use, we integrated session-level questionnaire ratings, TFO timing and use records, post-study SUS scores, and parent interviews. We organize the findings around routine integration, repeated-use acceptability, continued-use interest, use burden, and safety-related concerns.

% To understand how StoryEcho fit into families' daily routines and whether it was acceptable for repeated and continued home use, we analyzed parent interviews around routine integration, repeated-use acceptability, continued-use intention, use burden, and safety-related concerns.

\begin{figure*}[t]
  \includegraphics[width=\textwidth]{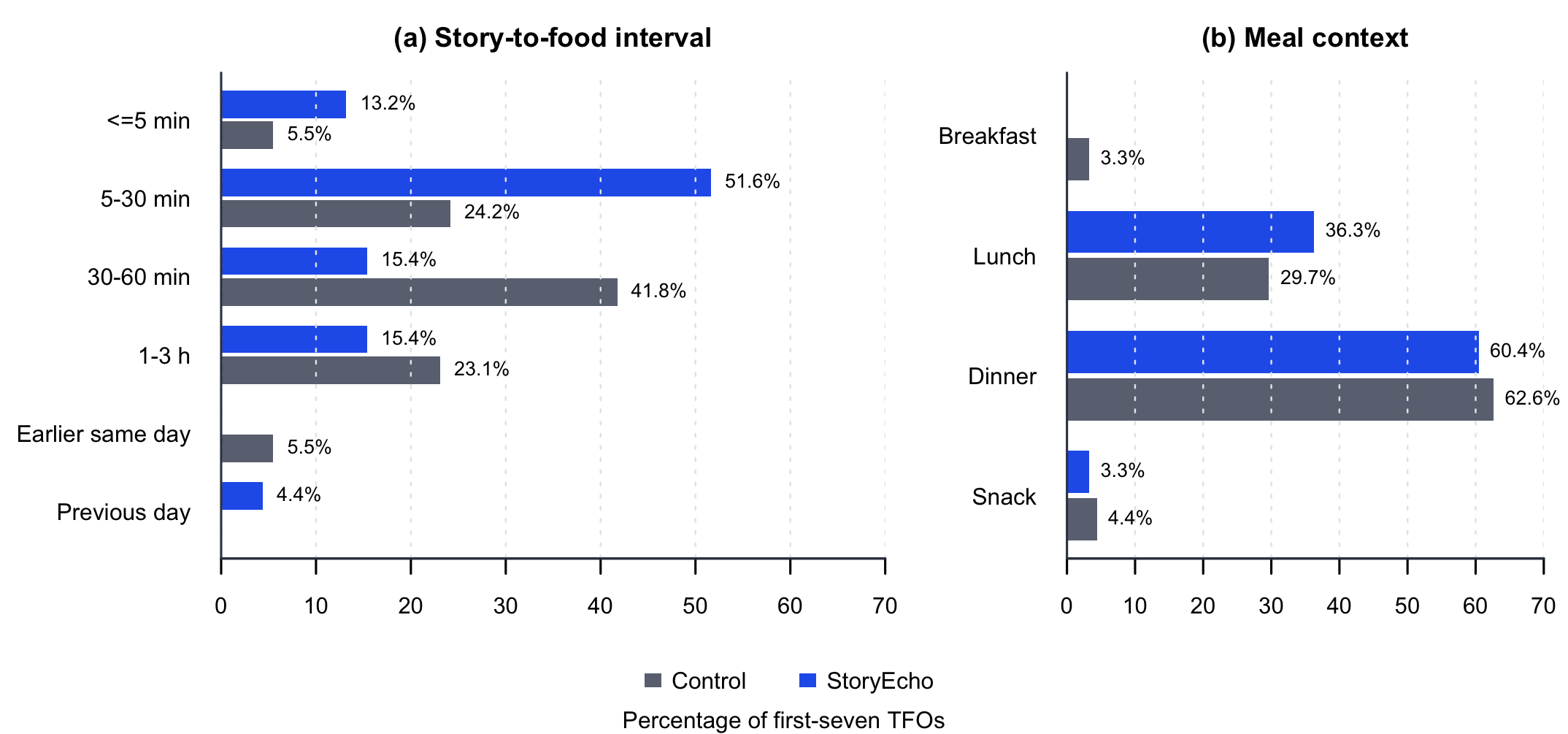}
  \caption{Timing and meal context of first seven target-food opportunities (TFOs) in the StoryEcho and Control conditions.}
  \label{fig:tfo_timing_context}
\end{figure*}

\textbf{Meal-adjacent but flexible routine integration.}
StoryEcho use was often organized around meals while remaining flexible. As shown in Fig.~\ref{fig:tfo_timing_context}, most target-food encounters occurred close to story use, especially within 5--30 minutes in the StoryEcho condition. Dinner was the most common context, likely because families had more control over evening meals (P1, P4, P7); P7 noted that they read after the child returned from kindergarten and before dinner at home. Families also embedded StoryEcho into ordinary household flows rather than treating it as a separate activity (P3, P5, P9, P11, P12, P13). For example, P13 generated the story before cooking, read it during cooking gaps, and let the continuation generate while washing dishes. Thus, StoryEcho was typically woven into meal preparation, eating, and post-meal cleanup as a flexible, meal-adjacent workflow.

\textbf{Higher enjoyment and growing acceptance of repeated use.}
Reported session ratings indicated that StoryEcho was more enjoyable than the Control condition. StoryEcho was associated with greater odds of higher \textit{child enjoyment} ($OR = 9.52$, $95\%~CI = [1.69, 53.70]$, $p_{\mathrm{Holm}} = .0426$), \textit{joint caregiver--child enjoyment} ($OR = 7.69$, $95\%~CI = [1.41, 41.97]$, $p_{\mathrm{Holm}} = .0426$), and \textit{continued-use intention} ratings ($OR = 10.82$, $95\%~CI = [1.22, 96.08]$, $p_{\mathrm{Holm}} = .0426$).

Interview accounts suggested that repeated-use acceptance developed as both caregivers and children became familiar with the routine. Parents described StoryEcho as easier to use after the first few sessions, as they learned how to time story generation, guide recording, and fit the loop into daily routines (P1, P4, P7). P1 summarized this as, \textit{``After doing it more times, it felt smoother.''} Children also began to actively anticipate StoryEcho as a recurring picture-book activity (P2, P4); P4 noted that after each session, the child would ask, \textit{``Mom, are we doing it again tomorrow?''} This growing familiarity also pointed to a need for variation over longer use: P12 noted that after several sessions, the four story frameworks felt \textit{``somewhat repetitive''}, and the child's interest declined in the final sessions.

\textbf{Continued-use interest extended beyond the required deployment.}
Continued-use interest appeared both behaviorally and in interviews. Three families (P7, P3, P10) exceeded the required seven sessions, completing 2, 4 and 4 additional sessions during deployment. When the required deployment period ended, children asked why there were no new stories or whether they could still listen later (P1, P2, P9); P9's child asked, \textit{``Can I still listen to this story picture book later?''} Caregivers also asked about future access (P3, P10, P11). P11 asked whether \textit{``this program could still be used later,''} and P3 noted that it might help friends' children as well, expressing interest in recommending it to them. Together, these behavioral and interview-based signals indicate continued-use interest beyond the study requirement.

\textbf{Low overall burden despite prototype-level friction.}
Caregivers reported lower workflow burden for StoryEcho than for the Control condition. StoryEcho was associated with lower odds of a higher \textit{system burden} rating ($OR = 0.088$, $95\%~CI = [0.013, 0.574]$, $p_{\mathrm{Holm}} = .0426$). It also received a mean SUS score of 86.92 ($SD = 7.30$), placing it in the ``excellent'' range on the adjective-based interpretation proposed by~\citet{bangor2008empirical}. Together, these results indicate that caregivers perceived StoryEcho as manageable and highly usable.

Interviews clarified that this favorable assessment coexisted with a few sources of prototype-level friction. Parents generally did not find StoryEcho's conceptual workflow difficult to understand; instead, reported burdens mostly concerned implementation issues such as generation latency, occasional technical delays, recording errors, and the effort of regenerating content after review. For example, P12 noted that the story had to be generated in advance because it was not ready \textit{``when I wanted to read it.''} These issues added planning and interaction costs, but parents tended to frame them as prototype limitations rather than reasons to reject the overall StoryEcho workflow.

\textbf{Caregiver oversight supported safety and low-pressure use.}
Parents generally did not report severe or sustained child distress during StoryEcho use. Caregiver-in-the-loop oversight supported safety: parents previewed content, co-read with children, explained tasks, and regenerated stories when needed. As P9 noted, \textit{``I looked through it every time; sometimes if I felt it was not good, I generated it again.''} Caregiver review identified two isolated content-production issues: an image depicting a child near a boiling pot that raised a potential safety concern (P10), and a sensory description inconsistent with the properties of winter melon (P12). These cases illustrate the value of caregiver oversight while pointing to opportunities to further improve visual safety and food-knowledge accuracy checks. Parents also valued StoryEcho's non-coercive framing (P1, P3, P4, P6, P8, P9, P13); P13 said it \textit{``did not force the child to try''} and that \textit{``if he really did not want to do it, that was also okay.''}

\section{Discussion}
% 多写一点

% 首先表明这个机制是effective（proved但是不需要写太多了），然后找一些心理学的理论解释一下为啥narrative mirroring loop有positive effect。

% interesting findings: 泛化性（整体积极性）？消除负面的情绪

% limitation+future work：一些原型使用的问题，表情可能的压力？小孩形象太单一？最后几页有时候没办法小孩没法理解？

\subsection{Why the Narrative Mirroring Loop May Support Low-Preference Food Exploration}

Our results suggest that the narrative mirroring loop helped children become more willing to engage with and try low-preference foods. Across repeated use, families reported increased food encounters, tries, and intake, alongside reduced caregiver pressure and gradually emerging child-initiated uptake.

Psychologically, the narrative mirroring loop may have worked by coupling two complementary mechanisms. 
The story-to-practice bridge aligns with Social Cognitive Theory~\cite{bandura1986social,bandura2001social}, which suggests that children can learn from symbolic models and by observing how actions lead to meaningful outcomes. In StoryEcho, the narrative counterpart functioned as such a symbolic model by demonstrating gradual sensory exploration of low-preference foods without imposing direct demands to eat. Because children selected a counterpart they found as similar or appealing, identification with the character may also have helped them imagine themselves performing similar actions within the story world~\cite{cohen2018defining}. The real-to-story mirroring complements this observational learning by reflecting children’s own experiences back into the narrative. This direction aligns with Self-Determination Theory (SDT) \cite{ryan2000self} and Cognitive Evaluation Theory (CET) \cite{deci1971effects}. SDT emphasizes autonomy, competence, and relatedness as conditions that support internalization and self-motivation~\cite{ryan2000self}, while CET suggests that informational, rather than controlling, feedback can foster intrinsic motivation~\cite{deci1971effects,deci2013intrinsic}. By reflecting children's food encounters through avatar changes, encouraging textual feedback, and subsequent story development, StoryEcho made their actions visible and consequential without framing them as demands or rewarding them through unrelated external incentives. This process may have strengthened children's sense of agency and competence, and may help explain the gradual emergence of child-initiated uptake.

More broadly, narrative mirroring can be understood as a recurring story--behavior--story connection in which narrative engagement prompts manageable real-world action and lived experiences become consequential to subsequent narrative development. This design pattern may inform other low-pressure behavior-change contexts that require repeated practice and caregiver scaffolding.

\subsection{From First-Contact Breakthroughs to Broader Spillover}
% 1. 对抵触情绪非常高，完全不愿意接触，但是家长并不清楚原因的食物，storyecho有时有意想不到的好效果。因为孩子完全不接触，所以无法认定是食物本身口感、味道孩子很厌恶，一般家长要求孩子尝试的方法对这种强抵触的食物往往无法起效。但使用了系统后，孩子突破了无法接触的门槛，开始能够接触这种食物，继而有时发现口感是可以接受的，因此，在这类食物上会呈现出从完全无法接受，直接到能够吃甚至不再讨厌的巨大变化，让一些家长印象很深刻，例如P2提到害怕茄子、P3青菜会吐掉，开始愿意尝试；P7西红柿不允许出现在碗里、P9豇豆一点都不碰，主动愿意吃。
% 2. 使用storyecho之后对非目标食物有时有泛化效果，一些家长提到对目标食物相似的食物也开始愿意尝试，例如P3的青菜到各种绿色蔬菜，P2的青椒到红彩椒，一种食物愿意尝试后，一些其他外观类似的食物也随之增加了尝试程度。一些家长提到随着storyecho融入家庭routine，孩子建立了storyecho和吃饭之间的关联，对吃饭本身的态度变得积极了，例如P3, P8, P13提到孩子在吃饭的context下主动提起storyecho，增加了整体的积极性。

Beyond the primary results, parents' accounts suggested two ways StoryEcho may have extended beyond the planned target-food pathway. First, StoryEcho appeared especially useful for foods that children strongly resisted but had rarely or never actually tried. In these cases, parents often could not tell whether rejection reflected actual sensory dislike or anticipated fear, disgust, or generalized refusal, because the child would not cross the initial threshold of contact. StoryEcho's low-pressure narrative and sensory tasks sometimes helped children cross this first-contact threshold. Once contact became possible, some children found the food more acceptable than expected, producing large and memorable shifts from complete refusal to touching, tasting, or eating. Parents described similar breakthroughs for eggplant (P2), leafy greens (P3), tomato (P7), and yardlong beans (P9), which their children had previously refused but later became willing to try, and in some cases, chose to eat.

Second, some families reported spillover effects beyond the specific target foods. Children sometimes became more receptive to foods that looked similar or belonged to the same broad category, such as moving from one leafy green to other green vegetables (P3), or from green pepper to red bell pepper (P2). These observations suggest that narrative mirroring may not only influence responses to a single food, but also foster broader associations that make similar foods seem more approachable. Parents further described spillover to the overall mealtime atmosphere. As StoryEcho became part of family routines, some children began referring to its stories during meals and connecting their eating experiences with the narrative world (P3, P8, P13). Although these effects were not prespecified outcomes, these observations suggest that the system may help children overcome strong resistance to particular foods while also fostering a more positive orientation toward food exploration in everyday meals.
% the system may support both breakthroughs with highly resisted foods and more generalized positive framing of food exploration in everyday meals.

\subsection{Limitations and Future Work}
% 一些原型使用的问题:生成时间长需要等待、出现内容问题的时候无法灵活单页重新生成、形象选择可选项略少有些单一。一些隐性的安全和效果思考：avatar表情变化非笑容时，虽然目前使用反馈未涉及压力，但可能带来隐性压力；故事结尾生成有时候没办法小孩没法理解和自身行为的直接关联，可能削弱行为和故事关联意识的作用。
% 针对这些提future work，另外把narrative mirroring loop扩展到更多场景

Our study has several limitations. At the prototype level, story generation sometimes required noticeable waiting time, which could interrupt parent--child interaction and reduce children's sustained engagement. The system also did not support flexible regeneration of a specific page or story segment, so minor content problems required broader regeneration and additional waiting time. In addition, the available avatar options and story frameworks were relatively limited, which may have constrained children's identification with the narrative counterpart and reduced variation across repeated sessions. The design also raises subtler questions about feedback and interpretation: although our findings did not suggest that avatar expression changes increased pressure, non-smiling or less positive expressions could potentially be experienced by some children as implicit evaluation. Behavior-informed endings also did not always make the connection between children's real-world actions and narrative changes sufficiently clear for preschool children, which may weaken awareness of the story--behavior link.

The study design further limits the strength of causal claims. The deployment involved a modest number of families over two weeks, and behavioral outcomes were caregiver-reported, which may introduce reporting or expectancy effects. Target foods were also not fully balanced across conditions. The small sample resulted in chance imbalances in the target foods represented in the analyzed TFOs. Although our models adjusted for the corresponding food-specific baselines, the observed effects should be interpreted as promising evidence rather than definitive causal estimates. Because the control condition removed the narrative mirroring loop as an integrated mechanism, the study cannot isolate the independent contribution of each loop component.

Future work should improve both the technical robustness and interactional sensitivity of generative storytelling systems for family-based food intervention. Faster generation and fine-grained editing or page-level regeneration could preserve interaction flow while allowing caregivers to correct inappropriate or confusing content without restarting the story. Expanding avatar customization and diversifying story frameworks may strengthen identification with the narrative counterpart while sustaining interest over longer-term use. Future designs should also examine how feedback can remain encouraging and informational without becoming evaluative or pressure-inducing, especially when visual expressions or progress indicators reflect children's behaviors. Behavior-informed continuation could also be made more transparent and developmentally appropriate by generating endings that more explicitly mirror the child's sensory exploration and its impact on the story world.

Beyond picky eating, future research could explore how the narrative mirroring loop may support other behavior-change contexts, such as children's health, hygiene, or learning routines, and whether similar mechanisms can be adapted for other populations, including adults and older adults.

\section{Conclusion}

In this paper, we presented StoryEcho, an LLM-based generative storytelling system that supports preschool children's exploration of low-preference foods through a narrative mirroring loop. By connecting story-based engagement with real-world sensory practice and reflecting children's food encounters back into subsequent story development, StoryEcho shifts picky-eating support beyond mealtime pressure toward lightweight, repeatable, and routine-compatible family interaction. In a two-week between-subjects field study with 26 families, children with StoryEcho showed greater willingness to try target foods, while caregiver pressure decreased and child-initiated uptake gradually emerged. StoryEcho also fit into everyday home routines and showed potential for repeated and continued use, suggesting the feasibility of the narrative mirroring loop as a lightweight family-based intervention approach. These findings highlight the potential of LLM-based generative storytelling as an intervention medium for everyday behavior-change goals, and position narrative mirroring as a promising design approach for storytelling systems that support children's participation while connecting narrative experience and real-world action over time.

% In this paper, we presented StoryEcho, a narrative mirroring loop generative storytelling system for picky-eating intervention in everyday family contexts. Through a formative study, we identified design considerations for low-pressure, family-centered intervention beyond mealtimes and translated them into StoryEcho's narrative mirroring loop, in which children are represented in the story and shape later story development through their real-world food-related behavior. Through our field study, we found that StoryEcho increased children's willingness to engage with low-preference foods, supported more constructive food-related parent-child interaction, and fit recurring family routines through low-pressure use outside mealtimes. Taken together, our findings suggest that picky-eating intervention may benefit from moving beyond mealtime-only support and immediate intake as the sole target of change. More broadly, they point to the promise of narrative mirroring loop generative systems, in which children's lived behavior becomes part of the generative process itself, as a design direction for behavior-related support in family settings. While our study was limited in scale and duration, it opens opportunities for future work on longer-term, more diverse, and more multimodal child-centered generative interventions.

\bibliographystyle{ACM-Reference-Format}
\bibliography{ref}

@inproceedings{chen2024more,
  title={More Than Just Limits: How Technology Can Support Parents in Regulating Children's Eating Behaviors at Family Mealtimes},
  author={Chen, Yang and Yen, Ching Chiuan},
  booktitle={Extended Abstracts of the CHI Conference on Human Factors in Computing Systems},
  pages={1--8},
  year={2024}
}

@article{goodman1961snowball,
  title={Snowball sampling},
  author={Goodman, Leo A},
  journal={The annals of mathematical statistics},
  pages={148--170},
  year={1961},
  publisher={JSTOR}
}

@article{wolstenholme2020childhood,
  title={Childhood fussy/picky eating behaviours: A systematic review and synthesis of qualitative studies},
  author={Wolstenholme, Hazel and Kelly, Colette and Hennessy, Marita and Heary, Caroline},
  journal={International Journal of Behavioral Nutrition and Physical Activity},
  volume={17},
  number={1},
  pages={2},
  year={2020},
  publisher={Springer}
}

@article{del2024risk,
  title={Risk Factors and consequences of food neophobia and pickiness in children and adolescents: a systematic review},
  author={Del Campo, Carmen and Bouzas, Cristina and Tur, Josep A},
  journal={Foods},
  volume={14},
  number={1},
  pages={69},
  year={2024},
  publisher={MDPI}
}

@article{lafraire2016food,
  title={Food rejections in children: Cognitive and social/environmental factors involved in food neophobia and picky/fussy eating behavior},
  author={Lafraire, J{\'e}r{\'e}mie and Rioux, Camille and Giboreau, Agn{\`e}s and Picard, Delphine},
  journal={Appetite},
  volume={96},
  pages={347--357},
  year={2016},
  publisher={Elsevier}
}

@article{saltzman2019associations,
  title={Associations between father availability, mealtime distractions and routines, and maternal feeding responsiveness: An observational study.},
  author={Saltzman, Jaclyn A and Musaad, Salma and Bost, Kelly K and McBride, Brent A and Fiese, Barbara H},
  journal={Journal of Family Psychology},
  volume={33},
  number={4},
  pages={465},
  year={2019},
  publisher={American Psychological Association}
}

@inproceedings{murnane2020designing,
  title={Designing ambient narrative-based interfaces to reflect and motivate physical activity},
  author={Murnane, Elizabeth L and Jiang, Xin and Kong, Anna and Park, Michelle and Shi, Weili and Soohoo, Connor and Vink, Luke and Xia, Iris and Yu, Xin and Yang-Sammataro, John and others},
  booktitle={Proceedings of the 2020 CHI Conference on Human Factors in Computing Systems},
  pages={1--14},
  year={2020}
}

@inproceedings{peng2018trip,
  title={A trip to the moon: Personalized animated movies for self-reflection},
  author={Peng, Fengjiao and LaBelle, Veronica Crista and Yue, Emily Christen and Picard, Rosalind W},
  booktitle={Proceedings of the 2018 CHI Conference on Human Factors in Computing Systems},
  pages={1--10},
  year={2018}
}

@article{dovey2008food,
  title={Food neophobia and ‘picky/fussy’eating in children: a review},
  author={Dovey, Terence M and Staples, Paul A and Gibson, E Leigh and Halford, Jason CG},
  journal={Appetite},
  volume={50},
  number={2-3},
  pages={181--193},
  year={2008},
  publisher={Elsevier}
}

@article{o2012repeated,
  title={Repeated exposure in a natural setting: A preschool intervention to increase vegetable consumption},
  author={O'Connell, Meghan L and Henderson, Kathryn E and Luedicke, Joerg and Schwartz, Marlene B},
  journal={Journal of the Academy of Nutrition and Dietetics},
  volume={112},
  number={2},
  pages={230--234},
  year={2012},
  publisher={Elsevier}
}

@article{nekitsing2019taste,
  title={Taste exposure increases intake and nutrition education increases willingness to try an unfamiliar vegetable in preschool children: a cluster randomized trial},
  author={Nekitsing, Chandani and Blundell-Birtill, Pam and Cockroft, Jennie E and Hetherington, Marion M},
  journal={Journal of the Academy of Nutrition and Dietetics},
  volume={119},
  number={12},
  pages={2004--2013},
  year={2019},
  publisher={Elsevier}
}

@article{weiser2026food,
  title={Food-related sensory activities for children in educational settings: a scoping review},
  author={Weiser, Hilde and Waling, Maria and Bohm, Ingela},
  journal={Appetite},
  volume={216},
  pages={108259},
  year={2026},
  publisher={Elsevier}
}

@article{braga2022nutrition,
  title={Nutrition education strategies to promote vegetable consumption in preschool children: the Veggies4myHeart project},
  author={Braga-Pontes, C{\'a}tia and Sim{\~o}es-Dias, Sara and Lages, Marlene and Guarino, Maria P and Gra{\c{c}}a, Pedro},
  journal={Public Health Nutrition},
  volume={25},
  number={4},
  pages={1061--1070},
  year={2022},
  publisher={Cambridge University Press}
}

@inproceedings{costa2023parental,
  title={Parental feeding practices and children’s eating behaviours: an overview of their complex relationship},
  author={Costa, Alexandra and Oliveira, Andreia},
  booktitle={Healthcare},
  volume={11},
  number={3},
  pages={400},
  year={2023},
  organization={MDPI}
}

@article{braun2006using,
  title={Using thematic analysis in psychology},
  author={Braun, Virginia and Clarke, Victoria},
  journal={Qualitative research in psychology},
  volume={3},
  number={2},
  pages={77--101},
  year={2006},
  publisher={Taylor \& Francis}
}

@article{o2020intercoder,
  title={Intercoder reliability in qualitative research: Debates and practical guidelines},
  author={O’Connor, Cliodhna and Joffe, Helene},
  journal={International journal of qualitative methods},
  volume={19},
  pages={1609406919899220},
  year={2020},
  publisher={SAGE Publications Sage CA: Los Angeles, CA}
}

@incollection{kadomura2013sensing,
  title={Sensing fork: Eating behavior detection utensil and mobile persuasive game},
  author={Kadomura, Azusa and Li, Cheng-Yuan and Chen, Yen-Chang and Tsukada, Koji and Siio, Itiro and Chu, Hao-hua},
  booktitle={CHI'13 Extended Abstracts on Human Factors in Computing Systems},
  pages={1551--1556},
  year={2013}
}

@article{jo2020mamas,
  title={MAMAS: supporting parent--child mealtime interactions using automated tracking and speech recognition},
  author={Jo, Eunkyung and Bang, Hyeonseok and Ryu, Myeonghan and Sung, Eun Jee and Leem, Sungmook and Hong, Hwajung},
  journal={Proceedings of the ACM on Human-Computer Interaction},
  volume={4},
  number={CSCW1},
  pages={1--32},
  year={2020},
  publisher={ACM New York, NY, USA}
}

@inproceedings{ganesh2014foodworks,
  title={FoodWorks: tackling fussy eating by digitally augmenting children's meals},
  author={Ganesh, Sangita and Marshall, Paul and Rogers, Yvonne and O'Hara, Kenton},
  booktitle={Proceedings of the 8th Nordic Conference on Human-Computer Interaction: Fun, Fast, Foundational},
  pages={147--156},
  year={2014}
}

@inproceedings{cai2024see,
  title={" See, Hear, Touch, Smell, and,... Eat!": Helping Children Self-Improve Their Food Literacy and Eating Behavior through a Tangible Multi-Sensory Puzzle Game},
  author={Cai, Xueyan and Jin, Kecheng and Shi, Shang and Huang, Shichao and Huang, Ouying and Wang, Xiaodong and Cheng, Jiahao and Lin, Weijia and Yao, Jiayu and Hu, Yuqi and others},
  booktitle={Proceedings of the 23rd Annual ACM Interaction Design and Children Conference},
  pages={270--281},
  year={2024}
}

@article{wang2025play,
  title={Play With Morphing Food: Supporting Children-Food Interaction With an Interactive Cooking Toolkit},
  author={Wang, Guanyun and Shao, Yilin and Feng, Boyu and Wang, Mengge and Zhou, Xiaojing and Yan, Yifan and Li, Zhengke and Yang, Yue and Zhu, Kuangqi and Wang, Yanan and others},
  journal={International Journal of Human--Computer Interaction},
  volume={41},
  number={15},
  pages={9731--9751},
  year={2025},
  publisher={Taylor \& Francis}
}

@article{coe2024exploring,
  title={Exploring the senses of taste with young children: Multisensory discoveries of food},
  author={Coe, Jennifer and Manera, Lorenzo and Fooladi, Erik C},
  journal={Food and Foodways},
  volume={32},
  number={1},
  pages={7--34},
  year={2024},
  publisher={Taylor \& Francis}
}

@article{hoppu2015impact,
  title={Impact of sensory-based food education in kindergarten on willingness to eat vegetables and berries},
  author={Hoppu, Ulla and Prinz, Mira and Ojansivu, Pauliina and Laaksonen, Oskar and Sandell, Mari A},
  journal={Food \& Nutrition Research},
  volume={59},
  number={1},
  pages={28795},
  year={2015},
  publisher={Taylor \& Francis}
}

@article{reverdy2008effect,
  title={Effect of sensory education on willingness to taste novel food in children},
  author={Reverdy, Caroline and Chesnel, Florence and Schlich, Pascal and K{\"o}ster, EP and Lange, Chris},
  journal={Appetite},
  volume={51},
  number={1},
  pages={156--165},
  year={2008},
  publisher={Elsevier}
}

@article{mahmoudizadeh2025interplay,
  title={The Interplay Between Picky Eating, Other Eating Behaviors, and Obesity Indicators in Preschool Children},
  author={Mahmoudizadeh, Melika and Babashahi, Mina and Rajabzadeh-Dehkordi, Milad and Nouri, Mehran and Faghih, Shiva},
  journal={Food Science \& Nutrition},
  volume={13},
  number={9},
  pages={e70967},
  year={2025},
  publisher={Wiley Online Library}
}

@article{trofholz2017parents,
  title={How parents describe picky eating and its impact on family meals: A qualitative analysis},
  author={Trofholz, Amanda C and Schulte, Anna K and Berge, Jerica M},
  journal={Appetite},
  volume={110},
  pages={36--43},
  year={2017},
  publisher={Elsevier}
}

@inproceedings{fries2019parental,
  title={Parental Feeding Practices and Associations with Children's Food Acceptance and Picky Eating.},
  author={Fries, Lisa R and Van der Horst, Klazine},
  booktitle={Nestle Nutrition Institute Workshop Series},
  volume={91},
  pages={31--39},
  year={2019}
}

@article{heath2014let,
  title={Let's look at leeks! Picture books increase toddlers' willingness to look at, taste and consume unfamiliar vegetables},
  author={Heath, Philippa and Houston-Price, Carmel and Kennedy, Orla B},
  journal={Frontiers in Psychology},
  volume={5},
  pages={191},
  year={2014},
  publisher={Frontiers Media SA}
}

@article{zahedi2025impact,
  title={The impact of a fairytale-like story on the food choices of preschool children},
  author={Zahedi, Anoushiravan and Katembu, Stephen and Sind, Sharon Michelle and Sommer, Undine and Kimamo, Charles and Sommer, Werner},
  journal={Appetite},
  volume={206},
  pages={107839},
  year={2025},
  publisher={Elsevier}
}

@inproceedings{garzotto2010interactive,
  title={Interactive storytelling for children},
  author={Garzotto, Franca and Paolini, Paolo and Sabiescu, Amalia},
  booktitle={Proceedings of the 9th international conference on interaction design and children},
  pages={356--359},
  year={2010}
}

@inproceedings{zhang2022storydrawer,
  title={StoryDrawer: a child--AI collaborative drawing system to support children's creative visual storytelling},
  author={Zhang, Chao and Yao, Cheng and Wu, Jiayi and Lin, Weijia and Liu, Lijuan and Yan, Ge and Ying, Fangtian},
  booktitle={Proceedings of the 2022 CHI conference on human factors in computing systems},
  pages={1--15},
  year={2022}
}

@article{cai2025child,
  title={Child-AI Co-Creation: A Review of the Current Research Landscape and a Proposal for Six Design Considerations},
  author={Cai, Zhenyao and Han, Ariel and Zhou, Xiaofei and Gazulla, Eva Durall and Peppler, Kylie},
  journal={Proceedings of the 24th Interaction Design and Children},
  pages={916--922},
  year={2025}
}

@inproceedings{xu2025accompany,
  title={Accompany Sleep: Using GenAI to Create Bedtime Stories for Mediating Parent-Child Relationships in LBC Families},
  author={Xu, Wenjie and Yu, Zhoutong and Liu, Yikun and Ying, Fangtian},
  booktitle={Proceedings of the 2025 CHI Conference on Human Factors in Computing Systems},
  pages={1--19},
  year={2025}
}

@inproceedings{chen2025characterizing,
  title={Characterizing llm-empowered personalized story reading and interaction for children: Insights from multi-stakeholder perspectives},
  author={Chen, Jiaju and Tang, Minglong and Lu, Yuxuan and Yao, Bingsheng and Fan, Elissa and Ma, Xiaojuan and Xu, Ying and Wang, Dakuo and Sun, Yuling and He, Liang},
  booktitle={Proceedings of the 2025 CHI Conference on Human Factors in Computing Systems},
  pages={1--24},
  year={2025}
}

@inproceedings{chen2025scenic,
  title={SCENIC: A Location-based System to Foster Cognitive Development in Children During Car Rides},
  author={Chen, Liuqing and Song, Yaxuan and Lyu, Ke and Xiao, Shuhong and Shen, Yilang and Sun, Lingyun},
  booktitle={Proceedings of the 38th Annual ACM Symposium on User Interface Software and Technology},
  pages={1--18},
  year={2025}
}

@article{brooke1996sus,
  title={SUS-A quick and dirty usability scale},
  author={Brooke, John and others},
  journal={Usability evaluation in industry},
  volume={189},
  number={194},
  pages={4--7},
  year={1996},
  publisher={London, England}
}

@book{shulevitz1997writing,
  title={Writing with Pictures: How To Write and Illustrate Children's Books.},
  author={Shulevitz, Uri},
  year={1997},
  publisher={ERIC}
}

@article{montag2015words,
  title={The words children hear: Picture books and the statistics for language learning},
  author={Montag, Jessica L and Jones, Michael N and Smith, Linda B},
  journal={Psychological science},
  volume={26},
  number={9},
  pages={1489--1496},
  year={2015},
  publisher={Sage Publications Sage CA: Los Angeles, CA}
}

@article{bangor2008empirical,
  title={An empirical evaluation of the system usability scale},
  author={Bangor, Aaron and Kortum, Philip T and Miller, James T},
  journal={Intl. Journal of Human--Computer Interaction},
  volume={24},
  number={6},
  pages={574--594},
  year={2008},
  publisher={Taylor \& Francis}
}

@article{richards2026storytelling,
  title={Storytelling as a health education intervention for children and adolescents: A scoping review},
  author={Richards, Jules and Ullman, Amanda and Plummer, Karin},
  journal={Journal of Pediatric Nursing},
  volume={90},
  pages={19--30},
  year={2026},
  publisher={Elsevier}
}

@misc{bandura1986social,
  title={Social foundations of thought and action: A social cognitive theory},
  author={Bandura, Albert},
  year={1986},
  publisher={Prentice-hall}
}

@article{bandura2001social,
  title={Social cognitive theory: An agentic perspective},
  author={Bandura, Albert},
  journal={Annual review of psychology},
  volume={52},
  number={1},
  pages={1--26},
  year={2001},
  publisher={Annual Reviews 4139 El Camino Way, PO Box 10139, Palo Alto, CA 94303-0139, USA}
}

@incollection{cohen2018defining,
  title={Defining identification: A theoretical look at the identification of audiences with media characters},
  author={Cohen, Jonathan},
  booktitle={Advances in foundational mass communication theories},
  pages={253--272},
  year={2018},
  publisher={Routledge}
}

@article{deci1971effects,
  title={Effects of externally mediated rewards on intrinsic motivation.},
  author={Deci, Edward L},
  journal={Journal of personality and Social Psychology},
  volume={18},
  number={1},
  pages={105},
  year={1971},
  publisher={American Psychological Association}
}

@book{deci2013intrinsic,
  title={Intrinsic motivation and self-determination in human behavior},
  author={Deci, Edward L and Ryan, Richard M},
  year={2013},
  publisher={Springer Science \& Business Media}
}

@article{ryan2000self,
  title={Self-determination theory and the facilitation of intrinsic motivation, social development, and well-being.},
  author={Ryan, Richard M and Deci, Edward L},
  journal={American psychologist},
  volume={55},
  number={1},
  pages={68},
  year={2000},
  publisher={American Psychological Association}
}

\newpage
\appendix
\section{Questionnaire Items and Scoring}
\label{app:questionnaire_measures}

Sections~\ref{sec:questionnaire_baseline}, \ref{sec:questionnaire_control}, and \ref{sec:questionnaire_storyecho} present English translations of the three Chinese caregiver questionnaires in their original order: the food-specific baseline form (11 items), the Control TFO form (20 items), and the StoryEcho TFO form (34 items). The prefixes B, C, and E identify the forms, and the numbers preserve the original item sequence. Caregivers in both conditions completed one baseline form per nominated food after allocation and before deployment. They completed the form corresponding to their assigned condition after each TFO. All child-related responses were caregiver reports rather than child self-reports. Only B3, B4, C20, and E34 were optional.

Each item reports its wording, response options, scoring direction, and analytical use. Shared questions are retained in both TFO subsections so that each questionnaire can be read independently. H1 concerns food-exploration outcomes; RQ1 concerns loop enactment and perception; RQ2 concerns routine fit, acceptability, and safety-related experiences. Items E19--E32 were administered only in StoryEcho. Their absence from Control is not a zero score. Numerical response anchors retain their original direction; recoding is identified separately. For categorical indicators, the specified response categories are counted descriptively rather than assumed to be equally spaced scores.

\paragraph{Composite outcomes and recoding.}
Let $x_j$ denote the raw response to item C$j$ or E$j$ within the same TFO. The two H1 secondary composites were $\mathit{approach}=(x_9+x_{10})/2$ and $\mathit{caregiver\ pressure}=[x_{12}+(8-x_{13})]/2$. Both range from 1 to 7 in increments of 0.5. Higher approach scores indicate greater independence and willingness; higher caregiver-pressure scores indicate more prompting and greater experienced stress. The RQ2 system-burden score was $8-x_{15}$, with higher values indicating greater caregiver-reported procedural burden. Intake responses were coded from 0 to 5 in their listed order. Other numbered outcomes retained their original direction. These derived outcomes were not additional questionnaire items.

\paragraph{Source notes.}
The highest frequency category in B5 is reproduced as ``more than 6 times''; the source form does not cover exactly six encounters. The first prompting anchor in B10 and C12 contains a Chinese typographical error. Its English interpretation follows the corresponding unambiguous anchor in E12. The ``cannot say'' option in E26 is retained as uncertainty, not as a pressure level. The post-study SUS was administered separately and is not included in these three questionnaires.

\subsection{Food-Specific Baseline Questionnaire}
\label{sec:questionnaire_baseline}

Caregivers in both conditions completed this questionnaire for each nominated food after allocation and before deployment. Items B3 and B4 were optional.

\begingroup
\small
\begin{enumerate}[
    label=\textbf{B\arabic*.},
    ref=B\arabic*,
    leftmargin=3.0em,
    labelsep=0.6em,
    itemsep=0.7\baselineskip,
    parsep=0.12\baselineskip,
    topsep=0.5\baselineskip,
    partopsep=0pt
]

\item \textbf{Participant ID.}
\label{item:questionnaire_B1}
Your participant ID.
\par\nobreak
Response options: Free text.
\par\nobreak
Scoring: Identifier; no score direction.
Analysis use: Record linkage.

\item \textbf{Target food.}
\label{item:questionnaire_B2}
Which target food is this baseline report about?
\par\nobreak
Response options: Free text.
\par\nobreak
Scoring: Food identifier; no score direction.
Analysis use: Child--food baseline matching.

\item \textbf{Food category.} (optional)
\label{item:questionnaire_B3}
Which category does this food belong to?
\par\nobreak
Response options: Vegetables; meat, eggs, and dairy; staple foods; soy products; fruit; mixed ingredients; other.
\par\nobreak
Scoring: Categorical; no score direction.
Analysis use: Food-composition description / sensitivity analysis.

\item \textbf{Usual preparation.} (optional)
\label{item:questionnaire_B4}
How do you usually prepare this food?
\par\nobreak
Response options: Free text.
\par\nobreak
Scoring: Descriptive text; no score direction.
Analysis use: Baseline context.

\item \textbf{Previous exposure.}
\label{item:questionnaire_B5}
Approximately how many times has the child encountered this food in the past month?
\par\nobreak
Response options: 0 times; 1--2 times; 3--5 times; more than 6 times.
\par\nobreak
Scoring: Ordered frequency categories; not a precise exposure count.
Analysis use: Baseline context.

\item \textbf{Baseline liking.}
\label{item:questionnaire_B6}
How much does the child usually like this food?
\par\nobreak
Response options: 1: Strong dislike; 2: Dislike; 3: Slight dislike; 4: Neutral; 5: Slight liking; 6: Liking; 7: Strong liking.
\par\nobreak
Scoring: Higher = greater liking. Foods with $B6\leq4$ were eligible for the target-food pool.
Analysis use: Food eligibility; baseline comparison.

\item \textbf{Baseline perceived difficulty.}
\label{item:questionnaire_B7}
How difficult do you think this food is for the child to accept?
\par\nobreak
Response options: 1: Very easy to accept; 2: Quite easy to accept; 3: Somewhat easy to accept; 4: Moderate; 5: Somewhat difficult; 6: Quite difficult; 7: Very difficult.
\par\nobreak
Scoring: Higher = greater perceived difficulty accepting the food.
Analysis use: Baseline comparison.

\item \textbf{Baseline try level.}
\label{item:questionnaire_B8}
What was the highest try level reached by the child during the most recent encounter with this food?
\par\nobreak
Response options: 1: Unwilling to approach, does not allow the food on the table, or clearly avoids it; 2: Allows the food on the table or plate without contact; 3: Willing to observe or move closer to smell the food; 4: Willing to touch, pick up, move, or play with the food; 5: Willing to bring the food to the mouth, lick it, or lightly touch it to the lips; 6: Bites the food but spits it out or does not swallow it; 7: Swallows approximately 1--2 bites; 8: Consumes some of the food but still needs encouragement; 9: Consumes a larger amount but still needs encouragement; 10: Voluntarily consumes the food at an almost normal level.
\par\nobreak
Scoring: Higher scores indicate a higher ordered food-exploration or consumption category. Matched to later TFOs for the same child--food pair.
Analysis use: H1 baseline adjustment.

\item \textbf{Baseline resistance.}
\label{item:questionnaire_B9}
How strongly did the child resist this food during the most recent encounter?
\par\nobreak
Response options: 1: No resistance; 2: Slight hesitation; 3: Some resistance, but can accept the food; 4: Refuses or delays; needs prompting or encouragement; 5: Strong refusal; needs repeated guidance; 6: Very strong resistance, with crying, tantrums, or avoidance; 7: Extreme resistance; unable to continue eating.
\par\nobreak
Scoring: Higher = stronger resistance.
Analysis use: H1 baseline adjustment; baseline comparison.

\item \textbf{Baseline caregiver prompting.}
\label{item:questionnaire_B10}
How much persuasion or prompting did you need to provide during the most recent encounter with this food?
\par\nobreak
Response options: 1: No prompting needed; the child accepts the food spontaneously; 2: Almost no prompting needed; 3: Occasional prompting is sufficient; 4: Some prompting needed; 5: Repeated prompting needed; 6: Continuous persuasion needed; 7: Extensive persuasion or prompting needed, with continued resistance.
\par\nobreak
Scoring: Higher = more prompting or persuasion.
Analysis use: H1 baseline adjustment; baseline comparison.

\item \textbf{Food-related concerns.}
\label{item:questionnaire_B11}
Has the child experienced an allergy, intolerance, difficulty swallowing or chewing, severe vomiting, or a similar problem with this food? If unsure, please describe a specific example.
\par\nobreak
Response options: Yes; no; I am unsure.
\par\nobreak
Scoring: Categorical safety information; no numerical score.
Analysis use: Food-specific safety information.

\end{enumerate}
\endgroup

\subsection{Control TFO Questionnaire}
\label{sec:questionnaire_control}

Caregivers in the Control condition completed this questionnaire after each TFO. Item C20 was optional.

\begingroup
\small
\begin{enumerate}[
    label=\textbf{C\arabic*.},
    ref=C\arabic*,
    leftmargin=3.0em,
    labelsep=0.6em,
    itemsep=0.7\baselineskip,
    parsep=0.12\baselineskip,
    topsep=0.5\baselineskip,
    partopsep=0pt
]

\item \textbf{Participant ID.}
\label{item:questionnaire_C1}
Your participant ID.
\par\nobreak
Response options: Free text.
\par\nobreak
Scoring: Identifier; no score direction.
Analysis use: Record linkage.

\item \textbf{TFO sequence.}
\label{item:questionnaire_C2}
What is the sequence number of this TFO cycle?
\par\nobreak
Response options: Free text.
\par\nobreak
Scoring: Reported sequence; analysis uses protocol-matched TFO position within each family.
Analysis use: Protocol order / repeated-use analysis.

\item \textbf{Target food.}
\label{item:questionnaire_C3}
What is the target food for this TFO?
\par\nobreak
Response options: Free text.
\par\nobreak
Scoring: Food identifier; no score direction.
Analysis use: Baseline matching / same-food exposure order.

\item \textbf{Preselected food.}
\label{item:questionnaire_C4}
Is this one of the target foods selected before the study?
\par\nobreak
Response options: Yes; no.
\par\nobreak
Scoring: Categorical yes/no; not a behavioral outcome.
Analysis use: Protocol documentation.

\item \textbf{Preparation change.}
\label{item:questionnaire_C5}
Has the preparation method for the target food changed from previous occasions during the study? If this is the first time preparing this food during the study, select the corresponding option and specify the preparation method.
\par\nobreak
Response options: No, the method is unchanged; yes, the method has changed; this is the first time preparing this food during the study.
\par\nobreak
Scoring: Categorical; no common high-to-low direction.
Analysis use: Context / preparation sensitivity analysis.

\item \textbf{Meal context.}
\label{item:questionnaire_C6}
In what setting was the target food tried during this TFO?
\par\nobreak
Response options: Breakfast; lunch; dinner; snack; an informal food-trying occasion.
\par\nobreak
Scoring: Categorical frequencies; no score direction.
Analysis use: RQ2 routine placement.

\item \textbf{Story--food interval.}
\label{item:questionnaire_C7}
How long before encountering the target food did the child use the story system?
\par\nobreak
Response options: Within 5 minutes; 5--30 minutes before; 30 minutes--1 hour before; 1--3 hours before; earlier the same day; the previous day; the child did not read a story in the system.
\par\nobreak
Scoring: Report the original time categories separately. No-story-use is not assigned an elapsed time.
Analysis use: RQ2 routine placement.

\item \textbf{Try level.}
\label{item:questionnaire_C8}
What was the highest try level reached by the child with the target food during this TFO?
\par\nobreak
Response options: 1: Unwilling to approach, does not allow the food on the table, or clearly avoids it; 2: Allows the food on the table or plate without contact; 3: Willing to observe or move closer to smell the food; 4: Willing to touch, pick up, move, or play with the food; 5: Willing to bring the food to the mouth, lick it, or lightly touch it to the lips; 6: Bites the food but spits it out or does not swallow it; 7: Swallows approximately 1--2 bites; 8: Consumes some of the food but still needs encouragement; 9: Consumes a larger amount but still needs encouragement; 10: Voluntarily consumes the food at an almost normal level.
\par\nobreak
Scoring: Higher scores indicate a higher ordered food-exploration or consumption category.
Analysis use: H1 primary outcome.

\item \textbf{Child independence.}
\label{item:questionnaire_C9}
How did the child mainly come to approach or try the target food during this TFO?
\par\nobreak
Response options: 1: Did not approach or try the food; 2: Tried only after continuous persuasion; 3: Tried after repeated reminders; 4: Tried after one reminder; 5: Approached or tried largely on their own initiative; 6: Spontaneously showed interest and tried the food; 7: Spontaneously requested to look at, smell, touch, or try the food.
\par\nobreak
Scoring: Higher = more independently initiated approach or trying; raw value retained.
Analysis use: Component of approach.

\item \textbf{Willingness.}
\label{item:questionnaire_C10}
How much curiosity or willingness to try the target food did the child show during this TFO?
\par\nobreak
Response options: 1: Very unwilling; 2: Quite unwilling; 3: Somewhat unwilling; 4: No clear willingness; neutral; 5: Somewhat willing or curious; 6: Quite willing; actively attended to the food; 7: Very willing; spontaneously proposed trying the food.
\par\nobreak
Scoring: Higher = greater curiosity or willingness; raw value retained.
Analysis use: Component of approach.

\item \textbf{Resistance.}
\label{item:questionnaire_C11}
How strongly did the child resist the target food during this TFO? Rate only the target food, not the entire meal.
\par\nobreak
Response options: 1: No resistance; 2: Slight hesitation; 3: Some resistance, but can accept the food; 4: Refuses or delays; needs prompting or encouragement; 5: Strong refusal; needs repeated guidance; 6: Very strong resistance, with crying, tantrums, or avoidance; 7: Extreme resistance; unable to continue eating.
\par\nobreak
Scoring: Higher scores indicate stronger resistance; raw values retained.
Analysis use: H1 secondary outcome.

\item \textbf{Caregiver prompting.}
\label{item:questionnaire_C12}
How much prompting, persuasion, or urging did you use to get the child to encounter or try the target food?
\par\nobreak
Response options: 1: No prompting needed; the child accepts the food spontaneously; 2: Almost no prompting needed; 3: Occasional prompting is sufficient; 4: Some prompting needed; 5: Repeated prompting needed; 6: Continuous persuasion needed; 7: Extensive persuasion or prompting needed, with continued resistance.
\par\nobreak
Scoring: Higher scores indicate more prompting; raw values enter the caregiver-pressure composite.
Analysis use: Component of caregiver pressure.

\item \textbf{Caregiver stress.}
\label{item:questionnaire_C13}
How much stress did you experience in relation to the target food during this TFO?
\par\nobreak
Response options: 1: Very high stress, with frequent conflict; 2: High stress, requiring continuous management; 3: Noticeable stress; 4: Moderate; 5: Low stress; 6: Quite relaxed; 7: Completely relaxed, with no stress.
\par\nobreak
Scoring: Raw higher = less stress. Analysis stress score $=8-x_{13}$; higher = greater stress. This reversed score, not the raw response, enters caregiver pressure.
Analysis use: Component of caregiver pressure.

\item \textbf{Intake.}
\label{item:questionnaire_C14}
Approximately how much of the target food did the child actually swallow? Bite count is distinct from the try level above.
\par\nobreak
Response options: None; one small bite; 2--3 bites; 4--5 bites; more than 5 bites; most or all of the food.
\par\nobreak
Scoring: Analysis codes: 0, 1, 2, 3, 4, and 5, respectively. Higher = a higher intake category; the codes are not literal bite counts.
Analysis use: H1 secondary outcome.

\item \textbf{System burden.}
\label{item:questionnaire_C15}
How burdensome was this procedure for you?
\par\nobreak
Response options: 1: Very troublesome and difficult to complete; 2: Quite troublesome; 3: Somewhat troublesome; 4: Moderate; 5: Quite convenient; 6: Very convenient; 7: Very easy and straightforward to complete.
\par\nobreak
Scoring: Raw higher = greater ease. Analysis score $=8-x_{15}$; higher = greater caregiver-reported procedural burden.
Analysis use: RQ2 burden.

\item \textbf{Child enjoyment.}
\label{item:questionnaire_C16}
How much did the child like the story this time?
\par\nobreak
Response options: 1: Strong dislike; 2: Moderate dislike; 3: Slight dislike; 4: Average; 5: Slight liking; 6: Moderate liking; 7: Strong liking.
\par\nobreak
Scoring: Higher = greater caregiver-reported liking; raw value retained.
Analysis use: RQ2 acceptability.

\item \textbf{Joint caregiver--child enjoyment.}
\label{item:questionnaire_C17}
How enjoyable was the experience of using the system together with the child?
\par\nobreak
Response options: 1: Very unpleasant; 2: Quite unpleasant; 3: Somewhat unpleasant; 4: Moderate; 5: Quite pleasant; 6: Very pleasant; 7: Extremely pleasant.
\par\nobreak
Scoring: Higher = a more enjoyable shared-use experience; raw value retained.
Analysis use: RQ2 acceptability.

\item \textbf{Continued-use intention.}
\label{item:questionnaire_C18}
Would you be willing to do this again next time?
\par\nobreak
Response options: 1: Completely unwilling; 2: Quite unwilling; 3: Somewhat unwilling; 4: Unsure; 5: Quite willing; 6: Very willing; 7: Extremely willing.
\par\nobreak
Scoring: Higher = greater willingness to repeat the procedure; raw value retained.
Analysis use: RQ2 continued-use interest.

\item \textbf{Special circumstances.}
\label{item:questionnaire_C19}
Were there any special circumstances that affected the child during this TFO?
\par\nobreak
Response options: The child was ill; the child was very sleepy; the meal was rushed for some reason; eating away from home; guests were present during the meal; the child was already very hungry or full; other; none.
\par\nobreak
Scoring: Context categories; no score direction.
Analysis use: Contextual interpretation / sensitivity analysis.

\item \textbf{Picture-book recording.} (optional)
\label{item:questionnaire_C20}
Please upload an audio recording of the picture-book use. If the upload fails, you may send it to the researcher via WeChat.
\par\nobreak
Response options: File upload (paste or drag the file into the upload area, or add a local file).
\par\nobreak
Scoring: Optional audio file; no questionnaire score.
Analysis use: Optional supplementary record.

\end{enumerate}
\endgroup

\subsection{StoryEcho TFO Questionnaire}
\label{sec:questionnaire_storyecho}

Caregivers in the StoryEcho condition completed this questionnaire after each TFO. Item E34 was optional. Items E19--E32 were administered only in this condition.

\begingroup
\small
\begin{enumerate}[
    label=\textbf{E\arabic*.},
    ref=E\arabic*,
    leftmargin=3.0em,
    labelsep=0.6em,
    itemsep=0.7\baselineskip,
    parsep=0.12\baselineskip,
    topsep=0.5\baselineskip,
    partopsep=0pt
]

\item \textbf{Participant ID.}
\label{item:questionnaire_E1}
Your participant ID.
\par\nobreak
Response options: Free text.
\par\nobreak
Scoring: Identifier; no score direction.
Analysis use: Record linkage.

\item \textbf{TFO sequence.}
\label{item:questionnaire_E2}
What is the sequence number of this TFO cycle?
\par\nobreak
Response options: Free text.
\par\nobreak
Scoring: Reported sequence; analysis uses protocol-matched TFO position within each family.
Analysis use: Protocol order / repeated-use analysis.

\item \textbf{Target food.}
\label{item:questionnaire_E3}
What is the target food for this TFO?
\par\nobreak
Response options: Free text.
\par\nobreak
Scoring: Food identifier; no score direction.
Analysis use: Baseline matching / same-food exposure order.

\item \textbf{Preselected food.}
\label{item:questionnaire_E4}
Is this one of the target foods selected before the study?
\par\nobreak
Response options: Yes; no.
\par\nobreak
Scoring: Categorical yes/no; not a behavioral outcome.
Analysis use: Protocol documentation.

\item \textbf{Preparation change.}
\label{item:questionnaire_E5}
Has the preparation method for the target food changed from previous occasions during the study? If this is the first time preparing this food during the study, select the corresponding option and specify the preparation method.
\par\nobreak
Response options: No, the method is unchanged; yes, the method has changed; this is the first time preparing this food during the study.
\par\nobreak
Scoring: Categorical; no common high-to-low direction.
Analysis use: Context / preparation sensitivity analysis.

\item \textbf{Meal context.}
\label{item:questionnaire_E6}
In what setting was the target food tried during this TFO?
\par\nobreak
Response options: Breakfast; lunch; dinner; snack; an informal food-trying occasion.
\par\nobreak
Scoring: Categorical frequencies; no score direction.
Analysis use: RQ2 routine placement.

\item \textbf{Story--food interval.}
\label{item:questionnaire_E7}
How long before encountering the target food did the child use the story system?
\par\nobreak
Response options: Within 5 minutes; 5--30 minutes before; 30 minutes--1 hour before; 1--3 hours before; earlier the same day; the previous day; the child did not read a story in the system.
\par\nobreak
Scoring: Report the original time categories separately. No-story-use is not assigned an elapsed time.
Analysis use: RQ2 routine placement.

\item \textbf{Try level.}
\label{item:questionnaire_E8}
What was the highest try level reached by the child with the target food during this TFO?
\par\nobreak
Response options: 1: Unwilling to approach, does not allow the food on the table, or clearly avoids it; 2: Allows the food on the table or plate without contact; 3: Willing to observe or move closer to smell the food; 4: Willing to touch, pick up, move, or play with the food; 5: Willing to bring the food to the mouth, lick it, or lightly touch it to the lips; 6: Bites the food but spits it out or does not swallow it; 7: Swallows approximately 1--2 bites; 8: Consumes some of the food but still needs encouragement; 9: Consumes a larger amount but still needs encouragement; 10: Voluntarily consumes the food at an almost normal level.
\par\nobreak
Scoring: Higher scores indicate a higher ordered food-exploration or consumption category.
Analysis use: H1 primary outcome.

\item \textbf{Child independence.}
\label{item:questionnaire_E9}
How did the child mainly come to approach or try the target food during this TFO?
\par\nobreak
Response options: 1: Did not approach or try the food; 2: Tried only after continuous persuasion; 3: Tried after repeated reminders; 4: Tried after one reminder; 5: Approached or tried largely on their own initiative; 6: Spontaneously showed interest and tried the food; 7: Spontaneously requested to look at, smell, touch, or try the food.
\par\nobreak
Scoring: Higher = more independently initiated approach or trying; raw value retained.
Analysis use: Component of approach.

\item \textbf{Willingness.}
\label{item:questionnaire_E10}
How much curiosity or willingness to try the target food did the child show during this TFO?
\par\nobreak
Response options: 1: Very unwilling; 2: Quite unwilling; 3: Somewhat unwilling; 4: No clear willingness; neutral; 5: Somewhat willing or curious; 6: Quite willing; actively attended to the food; 7: Very willing; spontaneously proposed trying the food.
\par\nobreak
Scoring: Higher = greater curiosity or willingness; raw value retained.
Analysis use: Component of approach.

\item \textbf{Resistance.}
\label{item:questionnaire_E11}
How strongly did the child resist the target food during this TFO? Rate only the target food, not the entire meal.
\par\nobreak
Response options: 1: No resistance; 2: Slight hesitation; 3: Some resistance, but can accept the food; 4: Refuses or delays; needs prompting or encouragement; 5: Strong refusal; needs repeated guidance; 6: Very strong resistance, with crying, tantrums, or avoidance; 7: Extreme resistance; unable to continue eating.
\par\nobreak
Scoring: Higher scores indicate stronger resistance; raw values retained.
Analysis use: H1 secondary outcome.

\item \textbf{Caregiver prompting.}
\label{item:questionnaire_E12}
How much prompting, persuasion, or urging did you use to get the child to encounter or try the target food?
\par\nobreak
Response options: 1: No prompting needed; the child accepts the food spontaneously; 2: Almost no prompting needed; 3: Occasional prompting is sufficient; 4: Some prompting needed; 5: Repeated prompting needed; 6: Continuous persuasion needed; 7: Extensive persuasion or prompting needed, with continued resistance.
\par\nobreak
Scoring: Higher scores indicate more prompting; raw values enter the caregiver-pressure composite.
Analysis use: Component of caregiver pressure.

\item \textbf{Caregiver stress.}
\label{item:questionnaire_E13}
How much stress did you experience in relation to the target food during this TFO?
\par\nobreak
Response options: 1: Very high stress, with frequent conflict; 2: High stress, requiring continuous management; 3: Noticeable stress; 4: Moderate; 5: Low stress; 6: Quite relaxed; 7: Completely relaxed, with no stress.
\par\nobreak
Scoring: Raw higher = less stress. Analysis stress score $=8-x_{13}$; higher = greater stress. This reversed score, not the raw response, enters caregiver pressure.
Analysis use: Component of caregiver pressure.

\item \textbf{Intake.}
\label{item:questionnaire_E14}
Approximately how much of the target food did the child actually swallow? Bite count is distinct from the try level above.
\par\nobreak
Response options: None; one small bite; 2--3 bites; 4--5 bites; more than 5 bites; most or all of the food.
\par\nobreak
Scoring: Analysis codes: 0, 1, 2, 3, 4, and 5, respectively. Higher = a higher intake category; the codes are not literal bite counts.
Analysis use: H1 secondary outcome.

\item \textbf{System burden.}
\label{item:questionnaire_E15}
How burdensome was this procedure for you?
\par\nobreak
Response options: 1: Very troublesome and difficult to complete; 2: Quite troublesome; 3: Somewhat troublesome; 4: Moderate; 5: Quite convenient; 6: Very convenient; 7: Very easy and straightforward to complete.
\par\nobreak
Scoring: Raw higher = greater ease. Analysis score $=8-x_{15}$; higher = greater caregiver-reported procedural burden.
Analysis use: RQ2 burden.

\item \textbf{Child enjoyment.}
\label{item:questionnaire_E16}
How much did the child like the story or activity this time?
\par\nobreak
Response options: 1: Strong dislike; 2: Moderate dislike; 3: Slight dislike; 4: Average; 5: Slight liking; 6: Moderate liking; 7: Strong liking.
\par\nobreak
Scoring: Higher = greater caregiver-reported liking; raw value retained.
Analysis use: RQ2 acceptability.

\item \textbf{Joint caregiver--child enjoyment.}
\label{item:questionnaire_E17}
How enjoyable was the experience of using the system together with the child?
\par\nobreak
Response options: 1: Very unpleasant; 2: Quite unpleasant; 3: Somewhat unpleasant; 4: Moderate; 5: Quite pleasant; 6: Very pleasant; 7: Extremely pleasant.
\par\nobreak
Scoring: Higher = a more enjoyable shared-use experience; raw value retained.
Analysis use: RQ2 acceptability.

\item \textbf{Continued-use intention.}
\label{item:questionnaire_E18}
Would you be willing to do this again next time?
\par\nobreak
Response options: 1: Completely unwilling; 2: Quite unwilling; 3: Somewhat unwilling; 4: Unsure; 5: Quite willing; 6: Very willing; 7: Extremely willing.
\par\nobreak
Scoring: Higher = greater willingness to repeat the procedure; raw value retained.
Analysis use: RQ2 continued-use interest.

\item \textbf{Cycle completion.}
\label{item:questionnaire_E19}
Was this TFO cycle fully completed?
\par\nobreak
Response options: Fully completed; partially completed (missing parts: ...); not completed.
\par\nobreak
Scoring: Reported as completion categories, not as a food-exploration score.
Analysis use: Protocol completion.

\item \textbf{Food-related attention.}
\label{item:questionnaire_E20}
While reading the story during this TFO, did the child spontaneously attend to or talk about the target food?
\par\nobreak
Response options: No; yes, but mainly with caregiver guidance; yes, the child spontaneously mentioned the target food, color, smell, story characters, plot, or related features.
\par\nobreak
Scoring: Any attention: either yes category. Spontaneous mention: the child-initiated category only.
Analysis use: RQ1 attention and child-initiated uptake.

\item \textbf{Story interaction.}
\label{item:questionnaire_E21}
Were the interactive tasks in the story completed during this TFO? These include imitation and interactive tasks within the picture-book pages.
\par\nobreak
Response options: No; partially completed; mostly completed.
\par\nobreak
Scoring: The reported mostly-completed indicator counts only the final response category.
Analysis use: RQ1 loop enactment.

\item \textbf{Sensory actions.}
\label{item:questionnaire_E22}
Which actions did the child complete? Select all that apply.
\par\nobreak
Response options: Looking; smelling; touching; licking; biting; describing color and shape; describing smell; describing mouthfeel; other.
\par\nobreak
Scoring: Multiple selections allowed. These options identify completed actions, not who initiated them.
Analysis use: Food-exploration action record.

\item \textbf{Sensory task completion.}
\label{item:questionnaire_E23}
Did the child complete the sensory exploration task introduced after the story during this TFO? This refers to the small task on the final picture-book page.
\par\nobreak
Response options: No; partially completed; mostly completed.
\par\nobreak
Scoring: At least partial completion: partially or mostly completed.
Analysis use: RQ1 loop enactment.

\item \textbf{Shared reflection / recording.}
\label{item:questionnaire_E24}
Who completed the post-meal recording step during this TFO?
\par\nobreak
Response options: The caregiver alone; the caregiver led, with limited child participation; the caregiver and child together; the child spontaneously described the experience while the caregiver helped record it; the recording stage was not completed (including recording, feedback, and story continuation).
\par\nobreak
Scoring: Any child involvement: the three categories involving the child, including limited participation. Active description: spontaneous child description with caregiver assistance only.
Analysis use: RQ1 participation and child-initiated uptake.

\item \textbf{Feedback exposure.}
\label{item:questionnaire_E25}
Did the child view or hear the avatar feedback?
\par\nobreak
Response options: Yes; no.
\par\nobreak
Scoring: Exposure indicator: yes. Viewing or hearing feedback does not itself measure a positive response.
Analysis use: RQ1 loop enactment.

\item \textbf{Perceived feedback pressure on the child.}
\label{item:questionnaire_E26}
Do you think this feedback placed pressure on the child?
\par\nobreak
Response options: No pressure at all; slight pressure; cannot say; some pressure; noticeable pressure.
\par\nobreak
Scoring: Retain the five named categories. The presence of stated pressure comprises slight, some, or noticeable pressure. Cannot say is an uncertainty response, not a midpoint or absence of pressure.
Analysis use: RQ2 safety / emotional pressure.

\item \textbf{Ending exposure.}
\label{item:questionnaire_E27}
Did the child read or reread the story ending updated based on behavior during this TFO?
\par\nobreak
Response options: Yes; no.
\par\nobreak
Scoring: Ending-read indicator: yes.
Analysis use: RQ1 loop enactment.

\item \textbf{Noticing changes.}
\label{item:questionnaire_E28}
Did the child notice a change in the story ending or avatar?
\par\nobreak
Response options: Did not notice; noticed but said nothing; spontaneously commented on the change.
\par\nobreak
Scoring: Any noticing: either noticing category. Spontaneous commentary: the final category only.
Analysis use: RQ1 attention and commentary.

\item \textbf{Behavior--story understanding.}
\label{item:questionnaire_E29}
Did the child understand that changes in the story ending or avatar were related to how they ate?
\par\nobreak
Response options: Showed no sign of understanding; may have understood a little; explicitly stated or demonstrated that ``my behavior affects the story, character, or ending.''
\par\nobreak
Scoring: At least possible recognition: either understanding category. Explicit evidence: the final category only. All judgments are caregiver-reported.
Analysis use: RQ1 caregiver-reported recognition.

\item \textbf{Anticipation.}
\label{item:questionnaire_E30}
Did the child look forward to the next story change?
\par\nobreak
Response options: No; a little; showed clear anticipation or spontaneously mentioned next time.
\par\nobreak
Scoring: Any anticipation: a little or the final category. Marked anticipation / spontaneous mention: the final category only; this combined response does not distinguish the two behaviors.
Analysis use: RQ1 anticipation.

\item \textbf{Caregiver review.}
\label{item:questionnaire_E31}
Was the content reviewed by a caregiver before being shown to the child?
\par\nobreak
Response options: Yes; no.
\par\nobreak
Scoring: Review indicator: yes; the question asks about review before child exposure.
Analysis use: RQ2 content oversight.

\item \textbf{Inappropriate content.}
\label{item:questionnaire_E32}
Did this story contain inappropriate content?
\par\nobreak
Response options: Yes, a clear problem, and the story was regenerated or use was stopped; yes, a minor problem that did not affect use; no.
\par\nobreak
Scoring: Report each category; any issue comprises either problem category. Regeneration or stopping is part of the first response, not a separate measured item.
Analysis use: RQ2 content safety.

\item \textbf{Special circumstances.}
\label{item:questionnaire_E33}
Were there any special circumstances that affected the child during this TFO?
\par\nobreak
Response options: The child was ill; the child was very sleepy; the meal was rushed for some reason; eating away from home; guests were present during the meal; the child was already very hungry or full; other; none.
\par\nobreak
Scoring: Context categories; no score direction.
Analysis use: Contextual interpretation / sensitivity analysis.

\item \textbf{Picture-book recording.} (optional)
\label{item:questionnaire_E34}
Please upload an audio recording of the picture-book use. If the upload fails, you may send it to the researcher via WeChat.
\par\nobreak
Response options: File upload (paste or drag the file into the upload area, or add a local file).
\par\nobreak
Scoring: Optional audio file; no questionnaire score.
Analysis use: Optional supplementary record.

\end{enumerate}
\endgroup

\end{document}